\documentstyle[12pt]{article}

\newcommand{\sect}[1]{\setcounter{equation}{0}\section{#1}
\vspace{-7mm}\indent}

\begin{document}

\topmargin -5mm 
\oddsidemargin 5mm

\begin{flushright} 
OU-HET 262 \\ 
April 1997 
\end{flushright}

\begin{center}
\begin{Large} 
{\bf Toward Second-Quantization of $D5$-Brane}\\ 
\end{Large}

\vspace{25pt}

\noindent 
Toshio NAKATSU 

\vspace{18pt}

\begin{small}
{\it Department of Physics,}\\ 
{\it Graduate School of Science, Osaka University,}\\ 
{\it Toyonaka, Osaka 560, JAPAN}
\end{small}

\vspace{25pt}

\underline{Abstract} 
\end{center}

\vspace{10pt}

\begin{small}
A framework of second-quantization of 
$D5$-branes is proposed. 
It is based on the study of topology of the 
moduli space of their low energy effective 
worldvolume theory. 
Among the topological cycles which resolve 
singularities caused by overlapping $D5$-branes, 
it is introduced those cycles which duals, 
constituting a subspace of cohomology group 
of the moduli space, 
turn out to define the Fock space of 
the second-quantized $D5$-branes. 
The second-quantized operators are given by 
creation and annihilation operators of those 
cycles or their duals. 
\end{small}

\newpage
\sect{Introduction and Summary} 

               The discovery of the Dirichlet branes 
($D$-branes) \cite{Polchinski} provides us a route to 
the quantization of solitonic objects \cite{Strominger} 
in string theory. 
The quantum fluctuations of these solitonic objects are 
described by open strings with one or both of their 
boundaries constrained on them, for which they are named 
Dirichlet branes.

                  Consider the Dirichlet five-branes 
($D5$-branes) in Type IIB theory 
with open strings having the $U(n)$ Chan-Paton factors 
or Type IIB theory under the background of $n$ coincident 
$D9$-branes. When there are $k$ coincident $D5$-branes at 
$x^6= \cdots =x^9=0$, their low energy effective worldvolume 
theory is given by a six-dimensional supersymmetric $U(k)$ 
gauge theory with global $U(n)$-symmetry 
\cite{Polchinski-Witten},\cite{Witten1}. 
In this article we investigate this six-dimensional theory 
including the Fayet-Iliopoulos $D$-terms, 
$\zeta_{\dot{A}\dot{B}}\int\!d^6\!x 
~Tr~D^{\dot{A}\dot{B}}$. 
The degenerate vacua ($or$ moduli space), 
which is denoted by ${\cal M}(k)$, are 
determined by the $D$-flat conditions. 
${\cal M}(k)$ is the moduli space of $k$ $D5$-branes 
with open strings having the $U(n)$ Chan-Paton factors. 
Regarding auxiliary $D$-fields as a hyperk\"ahler 
momentum map, 
the moduli space is a $4nk$-dimensional 
hyperk\"ahler manifold obtained by the 
hyperk\"ahler quotient construction \cite{HKLR}, 
\begin{eqnarray} 
\left.
{\cal M}(k)=
\mu_{\dot{A}\dot{B}}^{-1}
(\zeta_{\dot{A}\dot{B}}) 
\right/ U(k)~~~, 
\nonumber 
\end{eqnarray}
where 
$\mu_{\dot{A}\dot{B}}$ is the hyperk\"ahler momentum 
map of $U(k)$-symmetry 
(a global part of the $U(k)$ gauge symmetry).

     Due to the coupling with open string 
$D$-branes are identified with the BPS states which 
have the Ramond-Ramond charges. These BPS states are 
allowed to have the bound states which are marginally 
stable \cite{Witten2},\cite{Sen}.
For the case of $D5$-branes 
it is discussed in\cite{Harvey-Moore},\cite{FKN} the possibility of 
identifying these BPS states with cohomology 
elements of the moduli space of $D5$-branes and 
thereby the second-quantized five-branes are proposed.  
The purpose of this article is to provide a general 
framework of the second-quantization of $D5$-brane 
argued in \cite{FKN}.

     In the next section we study topology of the 
moduli space utilizing the techniques developed by 
Nakajima \cite{Nakajima}. 
Due to $D5$-branes the Lorentz group $SO(9,1)$ reduces, 
at least, to $SO(5,1)\times SO(4)$. 
$SO(4)$, which originates from the rotations in the 
four-dimensions $(x^6,\cdots,x^9)$, is a global symmetry 
of the worldvolume theory. 
To study the vacua it turns out useful to fix a complex 
structure of ${\cal M}(k)$. One can fix it by giving a 
complex structure of the four-dimensions. 
The hyperk\"ahler quotient acquires the form, 
\begin{eqnarray}
{\cal M}(k)= 
\frac{\mu_{{\bf C}}^{-1}(0) \cap 
           \mu_{{\bf R}}^{-1}(\eta)}
     {U(k)}~~~,
\nonumber 
\end{eqnarray}
where the hyperk\"ahler momentum map 
$\mu_{\dot{A}\dot{B}}$ is decomposed into 
$\mu_{{\bf C}}$, the complex (holomorphic) part, 
and 
$\mu_{{\bf R}}$, the real part. 
The three constants $\zeta_{\dot{A}\dot{B}}$ 
(coupled with the Fayet-Iliopoulos terms) 
are rotated equal to zero except the only 
one component, which is denoted by $\eta$. 
$\eta$ is assumed to be positive. 
In the case of $\eta$ being positive, the moduli space 
${\cal M}(k)$ turns out to be a smooth hyperk\"ahler 
manifold.

           In order to preserve this complex structure, 
the structure group $SO(4)$ 
of the four-dimensions reduces to $U(2)$. 
An abelian part of $U(2)$ rotates $(z_1,z_2)$, 
the holomorphic coordinates of ${\bf C}^2$, by 
phases and also acts on the moduli space ${\cal M}(k)$. 
Its fixed points in ${\cal M}(k)$ are related with 
overlapping $D5$-branes. 
Any fixed point turns out to become zero as 
$\eta \rightarrow 0$. 
``zero" is a singularity of the moduli space obtained by 
setting all the components $\zeta_{\dot{A}\dot{B}}$ 
being zero. 
This singularity is caused by $k$ $D5$-branes 
overlapping at the origin $(z_1,z_2)=(0,0)$. 
From the perspective of four-dimensional gauge theory one 
can say it corresponds to the small size limit of 
$k$ $SU(n)$-instantons sitting at the origin.

                It is shown that one can associate an 
appropriate set of $n$ Young tableaux with each fixed 
point. 
This correspondence is not one-to-one because of 
degeneracies of the fixed points. Strictly speaking, 
each set of $n$ Young tableaux satisfying conditions 
(\ref{Young tableau}) corresponds to a fixed submanifold 
of the moduli space. 
Taking the Morse theoretical viewpoint 
\cite{Nakajima} topology of ${\cal M}(k)$ is described 
from considerations on these fixed submanifolds. 
The action of the abelian part of $U(2)$ is hamiltonian 
and its momentum map can be regarded, 
with an appropriate combination, 
as a perfect Morse function on the moduli space. 
The Poincar\'e polynomial of the moduli space 
turns out to have form (\ref{Poincare Poly})
\begin{eqnarray}
&&P_t({\cal M}(k))=
\sum_{(\Gamma_1,\cdots,\Gamma_n)}
t^{2 \left\{ 
n(k-\sum_{j=1}^{n}l(\Gamma_j))
    +\sum_{i<j}l(\Gamma_i \backslash \Gamma_j) 
       -\sum_{i<j}\nu(\Gamma_i \backslash \Gamma_j) 
\right\} } 
P_t({\cal F}_{(\Gamma_1,\cdots, \Gamma_n)}) 
\nonumber 
\end{eqnarray} 
where the summation is performed with respect to 
those set of $n$ Young tableaux satisfying conditions 
(\ref{Young tableau}). 
To each set of $n$ Young tableaux 
$(\Gamma_1,\cdots,\Gamma_n)$, 
the corresponding fixed submanifold is denoted by 
${\cal F}_{(\Gamma_1,\cdots,\Gamma_n)}$.  
Note that $l(\Gamma_i),l(\Gamma_i/\Gamma_j)$ and  
 $\nu(\Gamma_i/\Gamma_j)$ are positive integers 
introduced in the text. 
In this expression of the Poincar\'e polynomial,
terms related with a given set of $n$ 
Young tableaux $(\Gamma_1,\cdots,\Gamma_n)$ 
can be read as  
\begin{eqnarray}
&&
t^{2 \left\{ 
       n(k-\sum_{j=1}^{n}l(\Gamma_j))
             +\sum_{i<j}l(\Gamma_i \backslash \Gamma_j) 
       -\sum_{i<j}\nu(\Gamma_i \backslash \Gamma_j) 
\right\} } 
P_t({\cal F}_{(\Gamma_1,\cdots, \Gamma_n)}) 
\nonumber \\ 
&& =
t^{2 \left\{ 
       n(k-\sum_{j=1}^{n}l(\Gamma_j))
             +\sum_{i<j}l(\Gamma_i \backslash \Gamma_j) 
       -\sum_{i<j}\nu(\Gamma_i \backslash \Gamma_j) 
\right\}
+~dim~{\cal F}_{(\Gamma_1,\cdots,\Gamma_n)} }
\left( 1 +O(t^{-2}) \right)~~.
\nonumber 
\end{eqnarray}
A cycle which gives the leading will be called the 
maximal dimensional cycle labelled by 
$(\Gamma_1,\cdots,\Gamma_n)$ and denoted by 
$C_{(\Gamma_1,\cdots,\Gamma_n)}$. 
Among these topological cycles,  
those labelled by $(\Gamma,\emptyset,\cdots,\emptyset)$ 
are studied in Section 3 from the viewpoint of $D5$-branes. 
$\Gamma$ is an arbitrary Young tableau of $k$ boxes. 
One can write it explicitly by 
$\Gamma=\mbox{$[k_1,\cdots,k_l]$}$ where 
$k_i$ $(1 \leq i \leq l)$ are non-decreasing 
positive integers satisfying $k_1+\cdots +k_l=k$. 
We begin Section 3 by examining the case of 
$\Gamma=\mbox{$[k]$}$. 
An realization of the cycle 
$C_{({\scriptsize \mbox{$[k]$}},
\emptyset,\cdots,\emptyset)}$ is presented.    
Any point of this cycle is shown to 
describe the vacuum of overlapping $k$ $D5$-branes 
which admits $2(nk-1)$ additional degrees of freedom. 
(For the explicit form presented in the text, 
$k$ $D5$-branes are degenerate at the origin.) 
These parameters give a parametrization of the cycle 
$C_{({\scriptsize \mbox{$[k]$}},
\emptyset,\cdots,\emptyset)}$. 
They turn out to disappear 
when $\eta$ goes to zero. 
It means that this topological cycle is added 
to resolve the aforementioned singularity of 
the moduli space at 
$\zeta_{\dot{A}\dot{B}}=0$. 
By changing the center-of-mass of $k$ $D5$-branes 
from the origin to an arbitrary point, say, $P$ 
in the four-dimensions  
one can construct a cycle which is same as 
$C_{({\scriptsize \mbox{$[k]$}}, 
\emptyset,\cdots,\emptyset)}$  
except that $k$ $D5$-branes are degenerate 
at $P$, not at the origin. 
In order to make the position of $k$ coincident 
$D5$-branes, this cycle will be denoted by 
$C_{{\scriptsize \mbox{$[k]$}}}(P)$. 
It is isomorphic to 
$C_{({\scriptsize \mbox{$[k]$}},
\emptyset,\cdots,\emptyset)}$.

      Nextly we consider vacua of $k=k_1+\cdots+k_l$ 
$D5$-branes ($k_1 \geq \cdots \geq k_l$) in which each 
$k_i$ pieces are degenerate at $P_i$. 
In particular we concentrate on the vacuum obtained by 
superposing the vacua of $k_i$ $D5$-branes which belong 
to $C_{{\scriptsize \mbox{$[k_i]$}}}(P_i)$. 
The additional degrees of freedom of each constituent 
are now regarded as parameters of the superposed vacuum. 
They constitute a $2(nk-l)$-dimensional cycle of 
${\cal M}(k)$, which is shown to be identified with 
the maximal cycle 
$C_{({\scriptsize \mbox{$[k_1,\cdots,k_l]$}},
\emptyset,\cdots,\emptyset)}$. 
In order to make the positions of overlapping 
$k$ $D5$-branes manifest 
this $2(nk-l)$-dimensional cycle will be denoted by 
$C_{{\scriptsize \mbox{$[k_1,\cdots,k_l]$}}}
(P_1,\cdots,P_l)$.

              In Section 4 we investigate on 
superposition of two vacua one of which is the 
vacuum of $k$ $D5$-branes belonging to 
$C_{{\scriptsize \mbox{$[k_1,\cdots,k_l]$}}}
(P_1,\cdots,P_l)$ 
while the other is a generic vacuum of 
$\tilde{k}$ $D5$-branes. 
This superposition defines an inclusion 
of 
$C_{{\scriptsize \mbox{$[k_1,\cdots,k_l]$}}}
(P_1,\cdots,P_l) \times {\cal M}(\tilde{k})$ 
to ${\cal M}(k+\tilde{k})$. 
Its image consists of vacua of $k+\tilde{k}$ 
$D5$-branes in which the configurations of $k$ pieces, 
regarded as vacua of $k$ $D5$-branes, belong to 
$C_{{\scriptsize \mbox{$[k_1,\cdots,k_l]$}}}
(P_1,\cdots,P_l)$. 
By letting the $l$-positions of the overlapping 
$k$ $D5$-branes free in the four-dimensions 
we obtain a noncompact submanifold of 
${\cal M}(k+\tilde{k})$. It is denoted by 
${\cal C}_{{\scriptsize \mbox{$[k_1,\cdots,k_l]$}}}$. 
For each $i$, considering the inclusion of 
$C_{{\scriptsize \mbox{$[k_i]$}}}(P_i)$
$\times {\cal M}(\tilde{k}+\sum_{j \neq i}^lk_j)$ 
to ${\cal M}(k+\tilde{k})$ and  
letting the position of the overlapping 
$k_i$ $D5$-branes free, we also obtain the 
noncompact submanifold 
${\cal C}_{{\scriptsize \mbox{$[k_i]$}}}$. 
The intersection of these submanifolds turns out to be 
${\cal C}_{{\scriptsize \mbox{$[k_1,\cdots,k_l]$}}}$ : 
\begin{eqnarray} 
\cap_{i=1}^l 
{\cal C}_{{\scriptsize \mbox{$[k_i]$}}} 
= 
{\cal C}_{{\scriptsize \mbox{$[k_1,\cdots,k_l]$}}}
~~~. 
\nonumber 
\end{eqnarray}

                 The Poincar\'e duals of 
the noncompact submanifolds 
${\cal C}_{{\scriptsize \mbox{$[k_1,\cdots,k_l]$}}}$ 
and 
${\cal C}_{{\scriptsize \mbox{$[k_i]$}}}$ 
($1 \leq i \leq l$) are introduced,  
which are denoted respectively by 
${\cal O}_{{\scriptsize \mbox{$[k_1,\cdots,k_l]$}}}$ 
and 
${\cal O}_{{\scriptsize \mbox{$[k_i]$}}}$. 
Their degrees ($or$ ghost numbers \cite{FKN}) 
are $2(nk-l)$ and $2(nk_i-1)$. 
It is argued in the text that they can be made 
depend only on the local data of the resolutions 
of the singularities caused by overlapping $D5$-branes.  
The above intersection formula can be rephrased as 
\begin{eqnarray}
{\cal O}_{{\scriptsize \mbox{$[k_1,\cdots,k_l]$}}} 
= 
\wedge_{i=1}^l
{\cal O}_{{\scriptsize \mbox{$[k_i]$}}}~~~. 
\nonumber 
\end{eqnarray}

                It is also possible to interpret 
${\cal O}_{{\scriptsize \mbox{$[k_1,\cdots,k_l]$}}}$ 
as an element of the cohomology group 
$H^*({\cal M}(k))$. 
It is dual to the cycle 
$C_{({\scriptsize \mbox{$[k_1,\cdots,k_l]$}},
\emptyset,\cdots,\emptyset)}$. 
With this identification we define the subspace of 
$H^*({\cal M}(k))$, 
\begin{eqnarray}
\bigoplus_{k_1+\cdots+k_l=k}
{\bf C}
{\cal O}_{{\scriptsize \mbox{$[k_1,\cdots,k_l]$}}}~~~.
\nonumber 
\end{eqnarray}
Due to the intersection formula it can be written as  
\begin{eqnarray}
\bigoplus_{k_1+\cdots +k_l=k}{\bf C}
{\cal O}_{k_1}\wedge \cdots \wedge 
{\cal O}_{k_l}~~~~~~~~~~
\left( {\cal O}_{k_i} 
\equiv {\cal O}_{{\scriptsize \mbox{$[k_i]$}}}
\right)~~~~~. 
\nonumber 
\end{eqnarray}
Therefore, 
as a physical Hilbert space 
of the topological field theory \cite{FKN}, 
this subspace of $H^*({\cal M}(k))$ admits 
to have a structure analogous to the Fock space. 
By considering the direct sum of the 
moduli spaces the Fock space structure is fully recovered. 
This extension will be reasonable from the viewpoint of 
$D5$-branes since the number of $D5$-branes does not suffer 
apriori any restriction. 
(The four-dimensions is not compactified.) 
So, finally we obtain the following subspace 
of $\oplus_k H^*({\cal M}(k))$ : 
\begin{eqnarray}
{\cal H}_{total}\equiv 
\bigoplus_l \bigoplus_{k_1,\cdots,k_l}
{\bf C}{\cal O}_{k_1}\wedge \cdots \wedge 
{\cal O}_{k_l}~~~~~. 
\nonumber 
\end{eqnarray}

                  Our proposal can be summarized 
as follows : 
${\cal H}_{total}$ is the Fock space of the 
second-quantized $D5$-branes which allow the 
$U(n)$ Chan-Paton factors, in which 
${\cal O}_m$ can be identified with a marginally 
stable bound state of $m$ $D5$-branes. 
The second-quantized operators are introduced 
as the creation and annihilation operators of 
these bound states.

\sect{Moduli Space of D5-Branes}

                  The goal of this section is 
to describe topology of the moduli space of 
D5-branes which is introduced as the degenerate vacua 
of their effective worldvolume theory. 
We study it utilizing the techniques developed 
by Nakajima \cite{Nakajima},\cite{Nakajima-Lec}.

              We start by giving a brief review on 
the effective worldvolume theory of D5-brane. 
Let us consider Type IIB theory with  
open strings having the $U(n)$ Chan-Paton factor 
($or$ Type IIB theory under the background 
of $n$ coincident D9-branes). 
When there is a D5-brane 
an open string can have either 
Neumann boundary with the (anti-) 
fundamental representation 
of $U(n)$ or 
Dirichlet boundary on the D5-brane. 
The Dirichlet boundary has the index of the fundamental 
representation of $U(k)$ when $k$ D5-branes overlap 
\cite{Witten2}. 
The combinations of these boundary conditions 
correspond to three different open string sectors : 
Neumann-Neumann (NN), 
Dirichlet-Dirichlet (DD) 
and Dirichlet-Neumann (DN). 
The quantization of DD and DN strings leads to the 
(first) quantization of five-brane 
\cite{Polchinski},\cite{Polchinski-Witten},\cite{Witten1}.

                 Suppose that there are $k$ 
coincident D5-branes at $x^6=\cdots =x^9=0$. 
Their low energy effective worldvolume theory 
can be described by massless modes of 
the DD and DN strings. 
It is a six-dimensional supersymmetric 
$U(k)$ gauge theory 
with global $U(n)$-symmetry 
\cite{Polchinski-Witten},\cite{Witten1}.

              There are two kinds of 
massless bosonic modes.  
$A_{\mu}$ $(\mu =0,\cdots,5)$ 
give a $U(k)$ gauge field on the worldvolume. 
$X^i$ $(i=6,\cdots,9)$ are scalar fields 
which belong to the adjoint representation 
of $U(k)$. 
Their $U(k)$ gauge transformations are 
\begin{eqnarray}
A_{\mu}(x) &\mapsto& 
g(x)A_{\mu}(x)g(x)^{-1}-i \partial_{\mu}g(x)g(x)^{-1}~~,
\nonumber \\ 
X^i(x) &\mapsto& 
g(x)X^i(x)g(x)^{-1}~~,
\end{eqnarray}
where $g(x)\in U(k)$. 
These fileds are invariant under the global 
$U(n)$-rotation  
since the DD string has no Neumann  boundary. 
The vacuum expectation values of $X^i$ become 
the collective coordinates of D5-branes. 
The $SO(4) \simeq SU(2)\times SU(2)_R$ which originates 
from the rotations in the four-dimensions 
$(x^6,\cdots,x^9)$ is a global symmetry group 
of the worldvolume theory. 
$SU(2)_R$ will be identified with 
the $SU(2)$ R-symmetry. 
In order to make it clear, 
it is convenient to rewrite $X^i$ as 
\begin{eqnarray}
X_{A \dot{A}}=X^i\sigma^i_{A\dot{A}}~~,
\end{eqnarray}
where 
$\sigma^i_{A\dot{A}} = 
(i\tau^1,i\tau^2,i\tau^3,{\bf 1}_2)$. 
$\tau^{1,2,3}$ are the Pauli matrices and 
${\bf 1}_2$ is a $2 \times 2$ identity matrix. 
$A$ $(=1,2)$ and 
$\dot{A}$ $(=\dot{1},\dot{2})$ are respectively   
the $SU(2)$ and $SU(2)_R$ indices. 
In the DN sector, 
there is a $SU(2)_R$ doublet complex scalar 
$H_{\dot{A}}$. 
Since the DN string has both Neumann and Dirichlet 
boundaries, each $H_{\dot{A}}$ transforms as 
$({\bf k},{\bf \bar{n}})$ representation 
under the action of $U(k) \times U(n)$ : 
\begin{eqnarray}
H_{\dot{A}}(x) 
\mapsto 
g(x)H_{\dot{A}}(x)h^{-1}~~, 
\end{eqnarray}
where $(g(x),h) \in U(k) \times U(n)$. 
$H_{\dot{A}}$ is represented by a 
$k \times n$ complex matrix 
while $A_{\mu}$ and $X^i$ are represented by 
$k \times k$ hermitian matrices.

                   The bosonic part of 
the effective action will be given by 
\begin{eqnarray}
S_{boson} &=& 
\int \!d^6\!x~ 
Tr \left\{ 
\frac{1}{4}F^{\mu \nu}F_{ \mu \nu} 
+\frac{1}{2}D^{\mu}X^{A\dot{A}}D_{\mu}X_{A\dot{A}}
+D^{\mu}H_{\dot{A}}D_{\mu}\bar{H}^{\dot{A}} \right. 
\nonumber \\ 
&& 
~~~~~~~~~~~~~~~~~~
\left. 
-\frac{1}{2}D^{\dot{A}\dot{B}}D_{\dot{A}\dot{B}}
+D^{\dot{A}\dot{B}}(X_{A \dot{A}}X^A_{~\dot{B}}
+H_{\dot{A}}\bar{H}_{\dot{B}}) \right\} ~~~~, 
\label{effective action}
\end{eqnarray}
where $D_{\mu}= \partial_{\mu}+iA_{\mu}$ and 
$\bar{H}_{\dot{A}}
= \epsilon_{\dot{A}\dot{B}}(H_{\dot{B}})^{\dagger}$.   
The degenerate vacua ($or$ moduli space) are determined 
by the $D$-flat conditions obtained from 
(\ref{effective action}),  
\begin{eqnarray}
X_{A (\dot{A}}X^A_{~\dot{B})}
+H_{(\dot{A}}\bar{H}_{\dot{B})}=0~~, 
\end{eqnarray}
which coincides with the ADHM equation \cite{ADHM} 
of $SU(n)$-instantons on ${\bf R}^4$.

           One may modify effective action 
(\ref{effective action}) by adding 
the Fayet-Iliopoulos $D$-terms, 
\begin{eqnarray}
S_{F.I.}=
-\zeta_{\dot{A}\dot{B}}\int \!d^6\!x 
~Tr~D^{\dot{A}\dot{B}}~~~,
\end{eqnarray}
which change the $D$-flat conditions to 
\begin{eqnarray}
X_{A (\dot{A}}X^A_{~\dot{B})}
+H_{(\dot{A}}\bar{H}_{\dot{B})}
=\zeta_{\dot{A}\dot{B}}{\bf 1}_k~~,  
\label{ADHM}
\end{eqnarray}
where ${\bf 1}_k$ is a $k \times k$ identity matrix.
Notice that equation (\ref{ADHM}) can be regarded 
as the ADHM equation \cite{Kronheimer-Nakajima} 
of $SU(n)$-instantons 
in the gravitational instanson background 
\footnote{It is discussed in 
\cite{Douglas-Moore} 
from the viewpoint of D-brane.}.

\subsection*{D5-Brane Vacua as Hyperk\"ahler Quotient}

             Let us introduce the hyperk\"{a}hler structure 
on the space of $X_{A\dot{A}}$ and $H_{\dot{A}}$. 
The triplet of symplectic forms are given by 
\begin{eqnarray}
\omega_{\dot{A}\dot{B}}=~ 
Tr~ \left\{\frac{1}{2}
dX_{A(\dot{A}}\wedge dX^A_{~\dot{B})} 
+dH_{(\dot{A}}\wedge d\bar{H}_{\dot{B})} \right\},  
\label{symplectic forms}
\end{eqnarray} 
where, 
since we are now considering the classical vacua, 
the field variables 
$X_{A \dot{A}}$ and $H_{\dot{A}}$ are 
$c$-number matrices. 
The hyperk\"{a}hler structure may be extractable 
from the expansion   
\begin{eqnarray}
\omega_{\dot{A}\dot{B}}
&=&
\omega_II_{\dot{A}\dot{B}}
+\omega_JJ_{\dot{A}\dot{B}}
+\omega_KK_{\dot{A}\dot{B}},
\label{omega-IJK}
\end{eqnarray}
where 
\footnote{
$\bar{\sigma}_{ij}^{~\dot{A}}~_{\dot{B}}
=\frac{1}{4}
(\bar{\sigma}_i\sigma_j
      -\bar{\sigma}_j\sigma_i)^{\dot{A}}~_{\dot{B}}$, 
where  $\bar{\sigma}_i=\sigma_i^{\dagger}$.
}
$I^{\dot{A}}_{~\dot{B}}
\equiv 2\bar{\sigma}_{12}^{~\dot{A}}~_{\dot{B}}$ ,  
$J^{\dot{A}}_{~\dot{B}}
\equiv 2\bar{\sigma}_{13}^{~\dot{A}}~_{\dot{B}}$ 
and   
$K^{\dot{A}}_{~\dot{B}}
\equiv 2\bar{\sigma}_{14}^{~\dot{A}}~_{\dot{B}}$
are the bases of $SU(2)_R$ which satisfy 
$I^2=J^2=K^2=-1$ and $IJ=-JI=K$.

             Under the global $U(k)$-symmetry (the global 
part of the $U(k)$ gauge symmetry of the worldvolume theory) 
$X_{A\dot{A}}$ and $H_{\dot{A}}$ transform adjointly 
and vectorialy. Their infinitesimal transforms are 
\begin{eqnarray} 
\delta X_{A\dot{A}}&=&
\mbox{$[$}\Omega, X_{A\dot{A}} \mbox{$]$}, 
\nonumber \\ 
\delta H_{\dot{A}}&=&
\Omega H_{\dot{A}},~~~~ 
\nonumber \\
\left(~ \delta \bar{H}_{\dot{A}} \right.
&=& 
\left. 
-\bar{H}_{\dot{A}}\Omega 
~\right), 
\label{u(k)-action}
\end{eqnarray}
where $\Omega \in u(k)$. 
($\Omega^{\dagger}=-\Omega$). 
These infinitesimal transforms define a vector field 
$\xi_{\Omega}$  
on the space of the field variables. 
Notice that symplectic forms (\ref{symplectic forms}) 
are invariant 
under $U(k)$-action (\ref{u(k)-action}). 
Moreover they satisfy 
\begin{eqnarray}
i_{\xi_{\Omega}}\omega_{\dot{A}\dot{B}}=
~Tr~ \left\{ \Omega 
d \mu_{\dot{A}\dot{B}} \right\} ,
\label{inner-product1}
\end{eqnarray}
where 
\begin{eqnarray}
\mu_{\dot{A}\dot{B}} 
\equiv 
X_{A(\dot{A}}X^A_{~\dot{B})}+
H_{(\dot{A}}\bar{H}_{\dot{B})}. 
\label{momentum map}
\end{eqnarray}
``$i_{\xi_{\Omega}}$" in (\ref{inner-product1}) 
means taking an inner product 
by the vector field 
$\xi_{\Omega}$.
Equation (\ref{inner-product1}) besides 
the invariance of $\omega_{\dot{A}\dot{B}}$ 
show that the $U(k)$-action is hamiltonian  
with respect to the symplectic structures  
$\omega_{I,J,K}$ and 
that its momentum map is given 
by 
$\mu_{\dot{A}\dot{B}}
\equiv \mu_II_{\dot{A}\dot{B}}+\mu_JJ_{\dot{A}\dot{B}}
+\mu_KK_{\dot{A}\dot{B}}$, where $\mu_{I,J,K}$ 
are $u(k)$ (or $u(k)^{*}$)-valued.

              The hyperk\"{a}hler momentum map 
$\mu_{\dot{A}\dot{B}}$ is directly related 
with the $D$-flat conditions. 
Adding the Fayet-Iliopoulos $D$-terms, 
$\zeta_{\dot{A}\dot{B}}\int d^6x Tr D^{\dot{A}\dot{B}}$ 
$(\zeta^{\dot{A}}_{~\dot{B}} \in su(2)_R)$, 
$ D$-flat conditions (\ref{ADHM}) can be written as 
\begin{eqnarray} 
\mu_{\dot{A}\dot{B}}=\zeta_{\dot{A}\dot{B}}{\bf 1}_k~~. 
\label{Dflat}
\end{eqnarray} 
The classical moduli space 
of the low energy effective worldvolume theory  
is now given by   
the hyperk\"ahler quotient \cite{HKLR}
\begin{eqnarray} 
{\cal M}(k)=
\left. \left. \left\{ 
~(X_{A\dot{A}},H_{\dot{A}})~ \right|~ 
\mu_{\dot{A}\dot{B}}=\zeta_{\dot{A}\dot{B}}{\bf 1}_k ~
\right\} \right/ U(k).  
\label{moduli} 
\end{eqnarray}
(\ref{moduli}) can be also regarded as the moduli 
space of $k$ D5-branes. 
As for the dimensionality, by simply counting up 
the degrees of freedom, it turns out to be 
\begin{eqnarray}
dim~{\cal M}(k) = 4nk~~. 
\end{eqnarray}

\subsection*{Complex Structue of D5-Brane Vacua}

              Due to D5-branes the Lorentz group 
$SO(9,1)$ reduces to $SO(5,1)\times SO(4)$. 
The $SO(4)~(\simeq SU(2) \times SU(2)_R)$ 
which originates from the rotations in the four-dimensions is a 
global symmetry group of the worldvolume theory. 
($SU(2)_R$ is identified with the $SU(2)~ R$-symmetry.) 
To investigate moduli space (\ref{moduli}) it turns out useful 
to fix a complex structure in the four-dimensions. 
Let us introduce the complex structure by 
\begin{eqnarray}
x_{A\dot{A}}=
\left(
\begin{array}{cc} 
z_1 & \bar{z}_2 \\ 
-z_2 & \bar{z}_1 
\end{array} \right). 
\label{cpx structure}
\end{eqnarray} 
To preserve it the structure group 
$SO(4) \simeq SU(2)\times SU(2)_R$ 
must be restricted to 
$U(2) \simeq SU(2) \times U(1)_R$. 
(\ref{cpx structure}) also fixes  
a complex structure of the space of field variables. 
One may introduce the complex $k \times k$ matrices 
$B_a (a=1,2)$ instead of the hermitian matrices $X^i$ 
\begin{eqnarray}
X_{A\dot{A}}=
\left(
\begin{array}{cc} 
B_1 & B_2^{\dagger} \\ 
-B_2 & B_1^{\dagger} 
\end{array} \right). 
\label{cpx structure2}
\end{eqnarray}

           With these complex matrices one can write down 
the symplectic forms $\omega^{\dot{A}}_{~\dot{B}}$ 
as the combination  of $(1,1)$ and $(2,0)~((0,2))$ forms 
\begin{eqnarray}
\omega^{\dot{A}}_{~\dot{B}}=
\omega_{{\bf R}}
\left( \begin{array}{cc} 
\frac{1}{2} & ~ \\ ~ & -\frac{1}{2} \end{array} \right)
+
\omega_{{\bf C}}
\left( \begin{array}{cc} 
~ & 1 \\ 0 & ~ \end{array} \right)
+
\bar{\omega}_{{\bf C}}
\left( \begin{array}{cc} 
~ & 0 \\ 1 & ~ \end{array} \right), 
\label{symplectic forms2}
\end{eqnarray}
where 
$\omega_{{\bf R}}$ and $\omega_{{\bf C}}$ 
($\bar{\omega}_{{\bf C}}$) are 
respectively $(1,1)$ and $(2,0)~((0,2))$ forms 
on the space of field variables. 
As regards $D$-flat conditions (\ref{Dflat}) 
they acquire the following form : 
\begin{eqnarray}
&& 
\mu_{{\bf C}}=0 ~, 
\label{ADHM1} \\
&& 
\mu_{{\bf R}}=\eta {\bf 1}_k~~,
\label{ADHM2} 
\end{eqnarray}
where $\mu_{{\bf C},{\bf R}}$ are given by 
\begin{eqnarray}
\mu_{{\bf C}} &=& 
\mbox{$[$}B_1,B_2 \mbox{$]$}+
2H_{\dot{1}}H_{\dot{2}}^{\dagger} , 
\nonumber 
\\ 
\mu_{{\bf R}}&=& 
\mbox{$[$}B_1,B_1^{\dagger}\mbox{$]$}
+\mbox{$[$}B_2,B_2^{\dagger}\mbox{$]$} 
+2H_{\dot{1}}H_{\dot{1}}^{\dagger}
-2H_{\dot{2}}H_{\dot{2}}^{\dagger}.
\end{eqnarray} 
$\mu_{{\bf R}}$ and $\mu_{{\bf C}}$ are 
the momentum maps of the $U(k)$-action 
constructed respectively 
from $\omega_{{\bf R}}$ and $\omega_{{\bf C}}$.
Notice that, by using the $SU(2)_R$-rotations, 
we have set in (\ref{ADHM1}) and (\ref{ADHM2}) 
the three constants $\zeta^{\dot{A}}_{~~\dot{B}}$ 
equal to zero except the only one component 
which is denoted by $\eta$. 
$\eta$ is assumed to be positive.

              Throughout this section 
it is convenient to regard the $k \times k$ matrices 
$B_a$ and the $k \times n$ matrices $H_{\dot{A}}$ 
as elements of $Hom (V,V)$ and $Hom (W,V)$ where 
$V$ and $W$ respectively denote 
${\bf C}^k$ and ${\bf C}^n$.

                            The positivity of $\eta$  
ensures the following stability \cite{Nakajima} 
for any solution of $D$-flat conditions 
(\ref{ADHM1},\ref{ADHM2}) : 
\begin{eqnarray} 
\bullet~~&& 
\mbox{Any subspace $S$ of $V$ which 
satisfies $B_a(S) \subset S$ and 
$H_{\dot{1}}(W) \subset S$} 
\nonumber \\  
&& 
\mbox{is equal to $V$.}  
\label{stability condition}
\end{eqnarray}
Let us derive stability condition 
(\ref{stability condition}) : 
Suppose (\ref{stability condition}) does not hold. 
Then there exists a subspace $S$ 
$(\neq \emptyset)$ of $V$ such that 
$B_a(S)\subset S$ and $H_{\dot{1}}(W) \subset S$. 
Let $S_{\perp}$ be the subspace of $V$ which 
is orthogonal to $S$. 
Notice that the actions of 
$B_a$ and $B_a^{\dagger}$ are 
closed respectively on $S$ and $S_{\perp}$. 
Let $B_a|_S$ and $B_a^{\dagger}|_{S_{\perp}}$ 
denote their restrictions on $S$ and $S_{\perp}$. 
$B_a$ will acquire the form  
\begin{eqnarray}
B_a=
\left( 
\begin{array}{cc}
B_a|_S & 
D_a \\ 
0 & 
\left(
B_a^{\dagger}|_{S_{\perp}}
\right)^{\dagger} 
\end{array}
\right)~~~~
\left( 
B_a^{\dagger}=
\left(
\begin{array}{cc}
\left(
B_a|_S 
\right)^{\dagger} & 
0 \\ 
D_a^{\dagger} & 
B_a^{\dagger}|_{S_{\perp}} 
\end{array}
\right) \right)~~.
\end{eqnarray}
Equation (\ref{ADHM2}), if one restrict it 
on the subspace $S_{\perp}$, 
can be written as  
\begin{eqnarray}
\sum_{a} \mbox{$[  
\left(B_a^{\dagger}|_{S_{\perp}}\right)^{\dagger},
B_a^{\dagger}|_{S_{\perp}}]$}
-\sum_a D_a^{\dagger}D_a 
-2H_{\dot{2}}|_{S_{\perp}}
\left(H_{\dot{2}}|_{S_{\perp}}\right)^{\dagger}
=
\eta {\bf 1}_{S_{\perp}}~~, 
\end{eqnarray}   
where $H_{\dot{A}}|_{S_{\perp}}$ are 
the projections of $H_{\dot{A}}$ onto 
$S_{\perp}$. 
($H_{\dot{1}}|_{S_{\perp}}=0$.) 
Taking the trace of this equation leads 
a cotradiction 
because we set $\eta >0$. 
Therefore $S=\emptyset$.

      At this stage it might be convenient to remark on 
the $U(k)$-quotient which appears in (\ref{moduli}). 
It originates in the $U(k)$ gauge symmetry of 
the worldvolume theory. 
Any fixed point of the $U(k)$-action, if it exists, 
cause a singularity in moduli space (\ref{moduli}). 
This singularity relates to symmetry enhancement 
of the theory. 
So it is important to ask whether such a fixed point 
does appear or not. 
Let $(B_a,H_{\dot{A}})$ be a fixed point of the 
$U(k)$-action. 
There exists $g$ $\in U(k)$ which satisfies  
$gB_ag^{-1}=B_a$ and $gH_{\dot{A}}=H_{\dot{A}}$. 
Notice that $(B_a,H_{\dot{A}})$ 
satisfies the stability condition. 
Namely any vector $v$ of $V$ can be written 
in the form, 
$v=f(B)H_{\dot{1}}(w)$, 
where $f(B)$ is an appropriate polynomial 
of $B_a$ and $w$ is an element of $W$. 
The action of $g$ on this vector can be evaluated as 
\begin{eqnarray}   
gv &=& gf(B)H_{\dot{1}}(w) 
\nonumber \\ 
   &=& f(gBg^{-1})gH_{\dot{1}}(w)~~,
\end{eqnarray} 
which turns out equal to $v$. 
$gv=v$. 
Since $v$ is an arbitrary vector of $V$, 
it shows $g=1$. 
Therefore, owing to stability condition 
(\ref{stability condition}), the $U(k)$-action 
is transitive. The $U(k)$-quotient gives 
a smooth hyperk\"ahler manifold.

\subsection*{Fixed Points of $T^2$-Action}

                         An abelian subgroup of 
the residual $U(2) \simeq SU(2)\times U(1)_R$ 
is interesting in the sense 
that its fixed points will be related with 
overlapping D5-branes.   
This $U(1) \times U(1)_R$ $(or~T^2)$ symmetry 
will rotate the complex coordinates $z_a$ 
of the four-dimensions by the phases  
\begin{eqnarray}
&&
z_1 \mapsto e^{i \phi}z_1, ~~~
\nonumber \\
&&
z_2 \mapsto e^{i \theta}z_2. 
\label{torus action}
\end{eqnarray} 
As regards $B_a$ and $H_{\dot{A}}$ 
the $T^2$-action has the 
forms 
\begin{eqnarray}
&& B_1 \mapsto e^{i \phi}B_1~,~~~~ 
    H_{\dot{1}} \mapsto e^{i \phi}H_{\dot{1}},     
\nonumber \\
&& B_2 \mapsto e^{i \theta}B_2~,~~~~ 
     H_{\dot{2}} \mapsto e^{-i \theta}H_{\dot{2}}.   
\label{torus action BH}
\end{eqnarray}
Notice that the transforms of 
$H=(H_{\dot{1}},H_{\dot{2}})$ are modified 
in (\ref{torus action BH}) 
by an right $U(n)$-action of 
the Chan-Paton symmetry.  
The $U(1)_R$-symmetry rotates $H$ into   
$(e^{i (\theta+\phi)/2}H_{\dot{1}}$, 
$e^{-i(\theta+\phi)/2}H_{\dot{2}})$. 
Multiplying it by $e^{-i(\theta-\phi)/2}$ 
$\in U(n)$, 
we can identify the transform with  
$(e^{i \phi}H_{\dot{1}},e^{-i \theta}H_{\dot{2}})$.

        From the adjoint action of $SU(2)_R$ on 
$\omega^{\dot{A}}_{~\dot{B}}$ one can find that 
$U(1)_R$, the abelian part, 
rotates the symplectic forms as follows :  
\begin{eqnarray}
\omega_{{\bf R}} \mapsto 
\omega_{{\bf R}}, ~~~~~
\omega_{{\bf C}} \mapsto 
e^{i(\theta+\phi)}\omega_{{\bf C}}.  
\end{eqnarray} 
Notice that $\omega_{{\bf R}}$ is invariant under 
the $U(1)_R$-action. 
In fact the $T^2$-action is hamiltonian 
with respect to $\omega_{{\bf R}}$. 
The corresponding k\"ahler momentum map 
$\mu_{\phi,\theta}$ are given by 
\begin{eqnarray}
\mu_{\phi}&=& 
Tr \{ B_1B_1^{\dagger}
       +2H_{\dot{1}}H_{\dot{1}}^{\dagger} \}, 
\nonumber \\ 
\mu_{\theta}&=& 
Tr \{ B_2B_2^{\dagger}
       +2H_{\dot{2}}H_{\dot{2}}^{\dagger} \}.  
\label{momentum map of T}
\end{eqnarray}

             The action of $U(1) \times U(1)_R$ 
preserves $D$-flat conditions (\ref{Dflat}) and 
also commutes with that of $U(k)$. 
It means that 
$U(1) \times U(1)_R$ can act on the moduli space 
${\cal M}(k)$. 
Let us give a closer look on the fixed points of 
this $T^2$-action. 
Notice that any two solutions of 
$D$-flat conditions (\ref{Dflat}) 
which differ from each other 
by the $U(k)$-action should be identified in 
moduli space (\ref{moduli}). 
Therefore, 
for any fixed point $(B_a,H_{\dot{A}})$  
$ \in {\cal M}(k) $ there exists a homomorphism 
$\gamma~$ $U(1) \times U(1)_R \longrightarrow U(k)$ 
which satisfies 
\begin{eqnarray}
e^{i \phi}B_1 &=& 
\gamma(\phi,\theta)B_1\gamma(\phi,\theta)^{-1},
\nonumber \\ 
e^{i \theta}B_2 &=& 
\gamma(\phi,\theta)B_2\gamma(\phi,\theta)^{-1},
\nonumber \\
e^{i \phi}H_{\dot{1}} &=& 
\gamma(\phi,\theta)H_{\dot{1}}, 
\nonumber \\  
e^{-i \theta}H_{\dot{2}} &=&
\gamma(\phi,\theta)H_{\dot{2}}. 
\label{fixed point}
\end{eqnarray} 
Since $\gamma(\phi, \theta) $ is diagonalizable, 
we may decompose $V$ into the sum of the eigenspaces 
of $\gamma(\phi,\theta)$ :  
\begin{eqnarray}
V=\bigoplus_{p,q \in {\bf Z}}V(p,q)~~~~,  
\label{decompo}
\end{eqnarray}
where $V(p,q)$ is an eigenspace of $\gamma$ with 
its eigenvalue equal to $e^{i(p \phi+q \theta)}$, 
that is, 
$\left. \gamma(\phi,\theta) \right|_{V(p,q)}$
$=e^{i(p \phi+q \theta)}{\bf 1}_{V(p,q)}$.

      It is important to note that 
these eigenvalues $e^{-i(p \phi+q \theta)}$ 
besides eigenspaces $V(p,q)$ 
are restricted by $D$-flat conditions (\ref{Dflat}), 
especially by the positivity of $\eta$. 
$B_a$ and $H_{\dot{A}}$ will be shown to satisfy 
the following properties : 
\begin{eqnarray}
\bullet
&&~ 
B_1(V(p,q))=V(p+1,q) ~~,~~B_2(V(p,q))=V(p,q+1)~~.  
\label{surjectivity of B} 
\\
\bullet
&&~
Im~H_{\dot{1}}=V(1,0)~~,~~H_{\dot{2}}=0~~.
\label{surjectivity of H}
\end{eqnarray}
And the allowed eigenvalues 
which appear in (\ref{decompo}) 
will be restricted to  
\begin{eqnarray}
      p \geq 1~,~~ q \geq 0  .
\label{allowed eigenvalues}
\end{eqnarray}
Therefore the decomposition has the form 
\begin{eqnarray}
V=\bigoplus_{p \geq 1, q \geq 0}V(p,q)~~~~. 
\label{decompo2}
\end{eqnarray}

            Let us derive these properties : 
i) 
We remark that relations (\ref{fixed point}) imply  
$B_1(V(p,q))\subseteq V(p+1,q)$,
$B_2(V(p,q))\subseteq V(p,q+1)$ 
as for $B_a$ and 
$H_{\dot{1}}(W)\subseteq V(1,0)$, 
$H_{\dot{2}}(W)\subseteq V(0,-1)$ 
as for $H_{\dot{A}}$. 
Because $(B_a, H_{\dot{A}})$ satisfies  
stability condition (\ref{stability condition}), 
eigenspaces $V(p,q)$ with 
$p \leq 0$ or $q \leq -1$ can not appear 
in (\ref{decompo}), 
which means (\ref{allowed eigenvalues}). 
In particular it implies $V(0,-1)=\emptyset$. 
Thus we obtain $H_{\dot{2}}=0$. 
ii) 
Nextly we will derive surjectivity 
(\ref{surjectivity of H}) of $H_{\dot{1}}$. 
Suppose there exists a vector $v$ of $V(1,0)$ 
such that $H_{\dot{1}}^{\dagger}v=0$. 
Notice that condition (\ref{allowed eigenvalues}) 
implies $V(0,0)=V(1,-1)=\emptyset$,  
which tells us $B_a^{\dagger}v=0$.
The mulitiplication of $v$ by $\eta$ 
may be evaluated using $D$-flat condition 
(\ref{ADHM2}),  
\begin{eqnarray}
\eta v 
&=& 
\left(
\sum_a
\mbox{$[$}B_a,B_a^{\dagger}\mbox{$]$} 
+2H_{\dot{1}}H_{\dot{1}}^{\dagger}
-2H_{\dot{2}}H_{\dot{2}}^{\dagger} 
\right)v 
\nonumber \\ 
&=&  
-\sum_{a}B_a^{\dagger}B_av ,
\end{eqnarray}
which gives  
\begin{eqnarray}
\eta |v|^2=-\sum_a|B_av|^2. 
\end{eqnarray}
This cause a contradiction 
except for the case of $v=0$ 
since the L.H.S. of the equation is non-negative 
while the R.H.S. is non-positive. Thus we obtain 
$v=0$. It means $Im H_{\dot{1}}=V(1,0)$. 
iii) 
As regards (\ref{surjectivity of B})  
we shall derive the first equation. 
Let us begin by considering the case of 
$q=0$. 
Suppose there exists a vector $v$ of $V(p+1,0)$ 
such that $B_1^{\dagger}v=0$. 
Since we can now assume $p \geq 1$,   
both $B_2^{\dagger}v$ and $H_{\dot{1}}^{\dagger}v$ 
vanish. 
The multiplication of $v$ by $\eta$ can be also 
evaluated using $D$-flat condition 
(\ref{ADHM2}). 
It gives us the equality, 
\begin{eqnarray}
\eta |v|^2 = -\sum_a|B_av|^2 , 
\end{eqnarray}   
which means $v=0$. 
Thus we obtain $B_1(V(p,0))=V(p+1,0)$. 
For the case of $q > 0$, 
by using the induction on $q$,  
one can derive the equation, 
$B_1(V(p,q))=V(p+1,q)$. 
One can also prove the second equation of  
(\ref{surjectivity of B}) 
by repeating the same argument as above.

           What is the physical or geometrical meaning 
of these fixed points of the $U(1)\times U(1)_R$-symmetry? 
To answer this question it is convenient to consider 
their behavior under the limit of $\eta$ being zero. 
Let $(B_a,H_{\dot{A}})$ be a fixed point of 
the $T^2$-action. 
Owing to properties (\ref{surjectivity of B}) 
we may regard  $B_a$ as upper trianglular matrices. 
$B_a=(b_{a~ij})$ where $b_{a~ij}=0$ for $i \geq j$. 
The $(i,i)$-component of equation (\ref{ADHM2}) can be 
written as follows :  
\begin{eqnarray}
\sum_a 
\left\{ 
\sum_{j=i+1}^k
|b_{a~ij}|^2
-\sum_{j=1}^{i-1} 
|b_{a~ji}|^2 
\right\} 
+2
\left( 
H_{\dot{1}}H_{\dot{1}}^{\dagger} 
\right)_{ii}
=\eta. 
\label{(i,i)-component}
\end{eqnarray} 
Let us consider an implication of equation 
(\ref{(i,i)-component}). 
For the case of $i=1$,  
setting $\eta=0$ in (\ref{(i,i)-component}),  
we obtain  
$b_{a~1j}=0$ $( 1 \leq ^{\forall} j \leq k)$ and 
$\left( 
H_{\dot{1}}H_{\dot{1}}^{\dagger} 
\right)_{11}=0$. 
For the case of $i \geq 2$, 
it is shown recursively that 
equations (\ref{(i,i)-component}) 
with $\eta$ being zero 
gives the conditions, 
$b_{a~ij}=0$ $( 1 \leq ^{\forall} j \leq k)$ and 
$\left( 
H_{\dot{1}}H_{\dot{1}}^{\dagger} 
\right)_{ii}=0$. 
Therefore we can conclude that 
{\it any fixed point $(B_a, H_{\dot{A}})$ 
goes to zero as $\eta \rightarrow +0$}. 
Notice that 
``zero" is a singularity of the moduli space 
at $\eta=0$. 
It is caused by $k$ D5-branes 
overlapping at the origin.  
From the perspective of four-dimensional gauge theory 
one can also say that it corresponds 
to the small size limit of $k$ $SU(n)$-instantons 
sitting at the origin. 
The above behavior of the fixed points shows 
that these fixed points and their associated cycles 
(which we will discuss in the next section) are those 
appearing by the resolution of the singularity which exists 
in the case of $\eta$ being zero.

           With each fixed point of the $T^2$-action 
it is possible to associate a set of $n$ Young tableaux 
by taking the following procedure. 
We first notice that, 
due to relation (\ref{surjectivity of B}), 
we can draw the following flow diagram 
among the eigenspaces 
$V(p,q)$ in (\ref{decompo2}) : 
\begin{eqnarray}
\begin{array}{ccccc}
\vdots   & ~           & \vdots     & ~  & ~
\\
\uparrow & ~           & \uparrow   & ~  & ~
\\ 
V(2,0)   & \rightarrow & V(2,1)     & \rightarrow & \cdots 
\\ 
\uparrow & ~           & \uparrow   & ~  & ~ 
\\ 
V(1,0)   & \rightarrow & V(1,1)     & \rightarrow & \cdots 
\end{array}
\end{eqnarray}
In this flow diagram the up- and right-arrows 
denote respectively the actions of $B_1$ and $B_2$. 
We also note that relations (\ref{surjectivity of B}) 
and (\ref{surjectivity of H}) give the inequalities 
on the dimensions of the eigenspaces in (\ref{decompo2})
\begin{eqnarray}
\bullet
&&~
dim~V(p,q) \geq dim~V(p\!+\!1,q)~~,~~ 
dim~V(p,q) \geq dim~V(p,q\!+\!1)~~. 
\nonumber \\ 
\bullet 
&&~
dim~V(p,q) \leq  n ~~(= dim~W)~~~. 
\label{dim inequality}
\end{eqnarray}
Due to these inequalities 
we may introduce $n$ subdiagrams among the eigenspaces. 
Let us draw a flow diagram among the eigenspaces 
which dimensions are not less than $d$ 
($1 \leq d \leq n$). 
(Note that we obtain the original diagram 
in the case of $d=1$.) 
For each $d$, 
replacing each eigenspace in the diagram by a box, 
we will obtain a Young tableau $\Gamma_d$. 
Dimensional inequalities (\ref{dim inequality}) 
can be rephrased as the conditions on 
these $n$ Young tableaux 
$\Gamma_1,\Gamma_2,\cdots,\Gamma_n$ :    
\begin{eqnarray}
\bullet&&~  
\Gamma_1 ~\supseteq~ \Gamma_2 ~\supseteq~
 \cdots ~\supseteq~ \Gamma_n~~.  
\nonumber \\
\bullet&&~
|\Gamma_1|+|\Gamma_2|+\cdots+|\Gamma_n|=k~~.   
\label{Young tableau}
\end{eqnarray} 
Here, 
for given two Young tableaux 
$\Gamma$ and $\tilde{\Gamma}$,  
we say 
$\Gamma \supseteq \tilde{\Gamma}$ 
when the Young tableau $\tilde{\Gamma}$ 
can be obtainable from 
the Young tableau $\Gamma$ 
by removing some boxes of $\Gamma$. 
The total number of boxes of $\Gamma$ 
is denoted by $|\Gamma|$.

         The correspondence between the fixed points 
of the $T^2$-action and 
the sets of the $n$ Young tableaux is not one-to-one. 
The fixed points will be degenerate. 
This degeneracy is due to the commutativity 
of the $U(1)\times U(1)_R$-symmetry and the 
$U(n)$ Chan-Paton symmetry. 
One can say that 
each set of $n$ Young tableaux which satisfy conditions 
(\ref{Young tableau}) corresponds to 
a fixed submanifold of the $T^2$-action.

\subsection*{Topology of D5-Brane Vacua}

        The fixed points of the $T^2$-action may be 
considered as the critical points of the following 
Morse function on the moduli space \cite{Nakajima}, 
\cite{Nakajima-Lec} 
\begin{eqnarray}
\mu_{T^2}\equiv\mu_{\phi}+\epsilon \mu_{\theta},  
\label{Morse function}
\end{eqnarray}
where $\epsilon (>0)$ is a perturbation parameter. 
Notice that the $T^2$-action is hamiltonian with respect to 
$\omega_{{\bf R}}$  which is regarded as 
the k\"ahler form of the moduli space.  
$\mu_{\phi,\theta}$ 
(\ref{momentum map of T}) are now the k\"ahler 
momentum map on the moduli space.

           Since the critical submanifolds of  
Morse function (\ref{Morse function}) are classified 
by the sets of $n$ Young tableaux 
$(\Gamma_1,\cdots,\Gamma_n)$ which satisfy 
conditions (\ref{Young tableau}), 
let us write the corresponding critical submanifold by 
${\cal F}_{(\Gamma_1,\cdots, \Gamma_n)}$. 
To describe the Morse indices besides the dimensions 
of these critical submanifolds, 
we shall prepare a few notations on Young tableaux : 
One can realize a Young tableau $\Gamma$ 
by a set of non-increasing positive integers, 
$\Gamma=\mbox{$[k_1,k_2,\cdots,k_l]$}$. 
($k_1 \geq k_2 \geq \cdots \geq k_l \geq 1$.) 
We introduce $l(\Gamma)$, the length of $\Gamma$, 
by $l(\Gamma)=l$. 
(Notice that the total number of boxes of $\Gamma$, 
which is denoted by $|\Gamma|$,  
is $k_1+k_2+\cdots +k_l$.)  
Let 
$\Gamma=\mbox{$[k_1,k_2,\cdots,k_l]$}$ 
and 
$\tilde{\Gamma}=
\mbox{$[\tilde{k}_1,\tilde{k}_2,\cdots,
\tilde{k}_{\tilde{l}}]$}$ 
be two Young tableaux which satisfy 
$\Gamma \supseteq \tilde{\Gamma}$. 
It means that $l \geq \tilde{l}$ and 
$k_i \geq \tilde{k}_i$ for $1 \leq i \leq \tilde{l}$. 
In such a situation it turns out useful  
to extend 
$\tilde{\Gamma}
=\mbox{$[\tilde{k}_1,\tilde{k}_2,
\cdots,\tilde{k}_{\tilde{l}}]$}$ by 
writting  
$\mbox{$[\tilde{k}_1,\tilde{k}_2,
\cdots,\tilde{k}_l]$}$ 
setting 
$\tilde{k}_i=0$ for $\tilde{l}+1 \leq i \leq l$.    
With this understanding 
let us introduce 
$l(\Gamma \setminus \tilde{\Gamma})$ as 
the number of $k_i$ 
$(1 \leq i \leq l)$ 
which satisfy $k_i > \tilde{k}_i$. 
And also we introduce 
$\nu(\Gamma \setminus \tilde{\Gamma})$ 
as the number of $\tilde{k}_i$ 
$(1 \leq i \leq l)$ 
which satisfy   
$\tilde{k}_i < k_i$ and 
$\tilde{k}_i=k_{i+1}$. 
($k_{l+1}\equiv 0$.)

           The dimensions of the critical submanifold 
${\cal F}_{(\Gamma_1,\cdots, \Gamma_n)}$  
turn out to be 
\begin{eqnarray}
dim~ {\cal F}_{(\Gamma_1,\cdots, \Gamma_n)}
=
2 \sum_{i<j}\nu (\Gamma_i \backslash \Gamma_j),   
\label{dim of F}
\end{eqnarray}
and the Morse index at 
${\cal F}_{(\Gamma_1,\cdots,\Gamma_n)}$ 
is given by 
\begin{eqnarray}
2 \left\{ 
n(k-\sum_{j=1}^{n}l(\Gamma_j))
    +\sum_{i<j}l(\Gamma_i \backslash \Gamma_j) 
       -\sum_{i<j}\nu(\Gamma_i \backslash \Gamma_j) \right\}.  
\label{Morse index}
\end{eqnarray}
Since the critical submanifolds have no odd dimensional 
cycles the Morse function is perfect and the Poincar\'e 
polynomial of the moduli space has the form 
\begin{eqnarray}
&&P_t({\cal M}(k))=
\sum_{(\Gamma_1,\cdots,\Gamma_n)}
t^{2 \left\{ 
n(k-\sum_{j=1}^{n}l(\Gamma_j))
    +\sum_{i<j}l(\Gamma_i \backslash \Gamma_j) 
       -\sum_{i<j}\nu(\Gamma_i \backslash \Gamma_j) 
\right\} } 
P_t({\cal F}_{(\Gamma_1,\cdots, \Gamma_n)}) 
\nonumber \\
&&~~~~~
\label{Poincare Poly}
\end{eqnarray}

                 Formulae (\ref{dim of F}) and 
(\ref{Morse index}) can be derived by using the 
techniques developed by Nakajima \cite{Nakajima}, 
\cite{Nakajima-Lec}. 
It is important to remark that 
one can also introduce \cite{Nakajima} moduli space 
(\ref{moduli}) by the complex symplectic quotient 
\begin{eqnarray} 
{\cal M}(k)=
\mu_{{\bf C}}^{-1}(0)^s / GL(k:{\bf C}), 
\label{moduli2}
\end{eqnarray}
where 
\begin{eqnarray}
\mu_{{\bf C}}^{-1}(0)^s =
\left\{ 
  (B_a,H_{\dot{A}}) 
\left| 
  \begin{array}{l} 
\cdot ~~
\mu_{{\bf C}}=0 \\ 
\cdot ~~
\mbox{stability condition (\ref{stability condition})} 
\end{array}
\right. \right\}. 
\end{eqnarray}
The action of $GL(k:{\bf C})$ in (\ref{moduli2}) 
is given by 
\begin{eqnarray}
&& 
B_a ~\mapsto~~gB_ag^{-1}~, 
\nonumber \\
&& 
H_{\dot{1}}~\mapsto~~gH_{\dot{1}}~,  
\nonumber \\
&& 
H_{\dot{2}}^{\dagger}~\mapsto
~~H_{\dot{2}}^{\dagger}g^{-1}~~,  
\label{GL(k:C)action}
\end{eqnarray}      
where $g \in GL(k:{\bf C})$. 
Roughly speaking, the quotient by $GL(k:{\bf C})$ 
means that we are considering not the unitary group but  
its complexfication as the gauge symmetry.

           Taking complex symplectic description 
(\ref{moduli2}) of the moduli space 
let us consider an infinitesimal deformation 
($B_a+\delta B_a,H_{\dot{A}}+\delta H_{\dot{A}}$) 
from a given vacuum ($B_a,H_{\dot{A}}$). 
In order that the infinitesimal deformation 
still describes a vacuum configuration,  
$(\delta B_a,\delta H_{\dot{A}})$ necessarily 
satisfies the equation, 
\begin{eqnarray} 
\mbox{$[\delta B_1,B_2]$}+
\mbox{$[B_1,\delta B_2]$}+ 
2\delta H_{\dot{1}} H_{\dot{2}}^{\dagger}+ 
2 H_{\dot{1}} \delta H_{\dot{2}}^{\dagger} 
=0 ~~.
\label{infinitesimal muC}
\end{eqnarray} 
The L.H.S. of (\ref{infinitesimal muC}) is nothing but 
the infinitesimal deviation of $\mu_{\bf C}$, 
which will be denoted by 
$\alpha(\delta B_a,\delta H_{\dot{1}},
\delta H_{\dot{2}}^{\dagger})$. 
Not every solution of equation 
(\ref{infinitesimal muC}) 
gives an independent vacuum. 
One should take account of gauge symmetry 
(\ref{GL(k:C)action}). The infinitesimal 
$GL(k:{\bf C})$-transform can be read as 
\begin{eqnarray} 
\delta_Y 
\left( \begin{array}{c}
B_a \\ 
H_{\dot{1}} \\ 
H_{\dot{2}}^{\dagger} 
\end{array} \right) 
= 
\left( \begin{array}{c} 
\mbox{$[Y,B_a]$} \\  
YH_{\dot{1}} \\  
-H_{\dot{2}}^{\dagger}Y  
\end{array}\right)~~~, 
\label{gl(k:C)action} 
\end{eqnarray} 
where $Y \in gl(k:{\bf C}) (\equiv Hom (V,V))$. 
The R.H.S. of equations (\ref{gl(k:C)action}) gives  
a matrix-valued function of $Y$, 
which will be denoted by $\beta(Y)$. 
Now the tangent space of the moduli space 
at a given vacuum 
$(B_a,H_{\dot{A}})$ can be realized 
using these $\alpha$ and $\beta$. 
Namely let us introduce the complex : 
\begin{eqnarray} 
Hom(V,V) 
\stackrel{\beta}{\longrightarrow} 
\begin{array}{c}
Hom(V,V)^{\oplus 2} \\ 
\oplus \\
Hom(W,V) \\ 
\oplus \\ 
Hom(V,W) 
\end{array} 
\stackrel{\alpha}{\longrightarrow} 
Hom(V,V)~~~, 
\label{tangent space} 
\end{eqnarray} 
where $\alpha \cdot \beta =0$ holds.  
Then we can identify 
the (holomorphic) tangent space at the vacuum 
($B_a,H_{\dot{A}}$) with $Ker~\alpha / Im~\beta$. 
Notice that, 
owing to the stability of $(B_a,H_{\dot{A}})$, 
$\alpha$ and $\beta$ in (\ref{tangent space}) 
turn out to be respectively 
surjective and injective.

        Let $(B_a,H_{\dot{A}})$ be a fixed point of 
the $U(1)\times U(1)_R$-symmetry. 
In such a case there exists a homomorphism 
$\gamma$ (cf.(\ref{fixed point}))  
and we can decompose $V$ 
into the direct sum of the eigenspaces of $\gamma$. 
(These eigenspaces have the corresponding 
$U(1)\times U(1)_R$-charges (cf.(\ref{decompo2})).) 
The spaces of matrices appearing in complex 
(\ref{tangent space}) can be also decomposed into  
the eigenspaces of $\gamma$ with the definite 
$U(1)\times U(1)_R$-charges 
\footnote{ 
$Hom (V,V)\simeq V^* \otimes V$ and  
$Hom (W,V)\simeq W^* \otimes V$.  
In these identifications 
we should count the $U(1)\times U(1)_R$-charges 
of the eigenspace $V(p,q)$ as $(-p,-q)$
(cf.(\ref{torus action tangent})).}. 
Since $\alpha$ and $\beta$ in (\ref{tangent space}) 
are respectively surjective and injective, 
it is possible  to obtain 
the character of $U(1)\times U(1)_R$ 
on the tangent space at the fixed point,   
by identifying it with 
$Ker~\alpha/Im~\beta$, 
from the characters on the spaces of matrices 
appearing in (\ref{tangent space}) 
\footnote{A little modification is needed since 
$\alpha$ and $\beta$ in (\ref{tangent space}) have 
the $U(1)\times U(1)_R$-charges. 
For the exact treatment we refer \cite{Nakajima}, 
in which $n=1$ case is studied.}.   
Counting the non-positive charges in the character  
on the tangent space we obtain formulae 
(\ref{dim of F}) and (\ref{Morse index}).

            To end this section 
it may be convenient to comment 
on the fixed submanifolds, 
especially those of the type 
$(\Gamma_1,\Gamma_2,\cdots,\Gamma_n)
=(\Gamma,\emptyset,\cdots,\emptyset)$.  
$\Gamma$ is a Young tableau of $k$ boxes.  
In these cases 
all the eigenspaces of $\gamma$ 
is one-dimensional, $dim V(p,q)=1$, 
and the rank of $H_{\dot{1}}$ is equal to one. 
From the corresponding flow diagrams one can normalize   
the $k \times n$ complex matrices $H_{\dot{A}}$ as 
\footnote{Here we use hyperk\"ahler description 
(\ref{moduli}) of the moduli space.}
\begin{eqnarray}
H_{\dot{1}}=
\left( 
\begin{array}{cccc}
0      & 0      & \ldots & 0 \\
\vdots & \vdots &        & \vdots \\ 
\vdots & \vdots &        & \vdots \\
0      & 0      & \ldots & 0 \\ 
\sqrt{k\eta/2}      & 0      & \ldots & 0  
\end{array} 
\right)~~,~~~~~H_{\dot{2}}=0. 
\label{H0}
\end{eqnarray}
Notice that the $U(n)$ Chan-Paton symmetry 
which commutes with the $U(1)\times U(1)_R$-symmetry 
can rotates $H_{\dot{A}}$ to $H_{\dot{A}}h$ 
where $h \in U(n)$. The critical submanifolds 
${\cal F}_{(\Gamma, \emptyset,\cdots, \emptyset)}$ 
will be generated by this $U(n)$-action. 
From the matrix forms of (\ref{H0}) one can see 
that the action of $U(n-1) \times U(1)$ (of $U(n)$) 
is irrelevant. (The action of $U(1)$, which 
gives the phase of $H_{\dot{1}}$,  
can be absorbed into the $U(k)$ gauge symmetry.) 
Therefore the critical submanifolds are  
\begin{eqnarray}
{\cal F}_{(\Gamma, \emptyset,\cdots, \emptyset)} &\simeq& 
\frac{U(n)}{U(n-1)\times U(1)}  
\nonumber \\ 
&=& {\bf C}P_{n-1}. 
\label{Pn-1}
\end{eqnarray}

\sect{Cycles of D5-Brane Vacua}

                 There exist several nontrivial 
cycles in the moduli space ${\cal M}(k)$. 
Apart from topologies of the critical submanifolds 
of Morse function (\ref{Morse function}) 
these nontrivial cycles will be labelled,  
first of all, by the sets of $n$ Young tableaux 
which satisfy condition (\ref{Young tableau}). 
In the Poincar\'{e} polynomial (\ref{Poincare Poly}),  
terms related with a given set of 
$n$ Young tableaux 
$(\Gamma_1,\cdots,\Gamma_n)$ are  
\begin{eqnarray}
&&
t^{2 \left\{ 
       n(k-\sum_{j=1}^{n}l(\Gamma_j))
             +\sum_{i<j}l(\Gamma_i \backslash \Gamma_j) 
       -\sum_{i<j}\nu(\Gamma_i \backslash \Gamma_j) 
\right\} } 
P_t({\cal F}_{(\Gamma_1,\cdots, \Gamma_n)}) 
\nonumber \\ 
&& =
t^{2 \left\{ 
       n(k-\sum_{j=1}^{n}l(\Gamma_j))
             +\sum_{i<j}l(\Gamma_i \backslash \Gamma_j) 
       -\sum_{i<j}\nu(\Gamma_i \backslash \Gamma_j) 
\right\}
+~dim~{\cal F}_{(\Gamma_1,\cdots,\Gamma_n)} }
\left( 1 +O(t^{-2}) \right)~~.
\nonumber \\ 
&&~~~ 
\label{expansion}
\end{eqnarray}
A cycle which gives the leading of 
(\ref{expansion}) will be called the maximal 
dimensional cycle labelled by 
$(\Gamma_1,\cdots,\Gamma_n)$. 
It will be denoted by 
$C_{(\Gamma_1,\cdots,\Gamma_n)}$. 
The goal of this section is the interpretation of 
these topological cycles in terms of D5-branes. 
In particular, 
we will investigate the maximal dimensional cycles 
which are labelled by 
$(\Gamma_1,\Gamma_2,\cdots,\Gamma_n)=$ 
$(\Gamma,\emptyset,\cdots,\emptyset)$.

             In order to proceed further we will need 
explicit forms of these cycles. 
These explicit forms may be handled by using 
another equivalent description (\ref{moduli2}) 
of the moduli space.

            Let us start by giving a simple 
example of these maximal cycles. 
Consider the case of $l(\Gamma)=1$.  
Namely $\Gamma = \mbox{$[k]$}$. 
The dimensions of the maximal cycle 
$C_{({\scriptsize \mbox{$[k]$}},
\emptyset,\cdots,\emptyset)}$ 
can be read from  
the dimensions of the critical submanifold 
${\cal F}_{({\scriptsize \mbox{$[k]$}},
\emptyset,\cdots,\emptyset)}$ 
and the corresponding Morse index 
\begin{eqnarray}
dim~ C_{({\scriptsize \mbox{$[k]$}},
\emptyset,\cdots,\emptyset)} &=& 
2(n-1)+2n(k-1) 
\nonumber \\ 
&=&
2nk-2 .
\label{dim of C1}
\end{eqnarray}
Taking complex symplectic description 
(\ref{moduli2}) of the moduli space,  
an explicit form of the vacuum which belongs 
to this maximal cycle will be given by 
\footnote{
The configuration given in (\ref{cycle1}) 
is a representative of the vacuum. 
Any configuration which can be obtained from 
(\ref{cycle1}) by $GL(k:{\bf C})$-action 
(\ref{GL(k:C)action}) describes the same vacuum.}
\begin{eqnarray}
&&
B_1 = 
\left( 
\begin{array}{ccccc}
0       & 1      & 0       & \ldots & 0        \\
0       & 0      & 1       & \ddots & \vdots   \\ 
\vdots  & \vdots & \ddots  & \ddots & 0        \\
\vdots  & \vdots &         & \ddots & 1        \\ 
0       &  0     & \ldots  & \ldots & 0 
\end{array} 
\right)
~~,~~  
B_2 = 
\left( 
\begin{array}{ccccc}
0       & a_1    & a_2     & \ldots  & a_{k\!-\!1}  \\
\vdots  & 0      & a_1     & \ddots  & \vdots   \\ 
\vdots  & \vdots & \ddots  & \ddots  & a_2      \\
\vdots  & \vdots &         & \ddots  & a_1        \\ 
0       & 0      & \ldots  & \ldots  & 0 
\end{array}
\right) 
\nonumber 
\\ 
&&
H_{\dot{1}} =
\left( 
\begin{array}{cccc}
0      & h_{11}    & \ldots & h_{1 n\!-\!1} \\
\vdots & \vdots    &        & \vdots \\ 
\vdots & \vdots    &        & \vdots \\
0      & h_{k\!-\!1 1} & \ldots & h_{k\!-\!1 n\!-1} \\ 
1      & h_{k 1}   & \ldots & h_{k n\!-\!1}
\end{array} 
\right) 
~~, ~~
H_{\dot{2}}=0~~. 
\label{cycle1}
\end{eqnarray} 
$a_i$ and $h_{ij}$ in (\ref{cycle1}) 
are the $nk-1$ complex parameters 
which constitute the maximal cycle 
$C_{({\scriptsize \mbox{$[k]$}},
\emptyset,\cdots,\emptyset)}$. 
Explicit form (\ref{cycle1}) 
may be obtained as follows. 
Let $(B_a,H_{\dot{A}})$ be 
a fixed point of $T^2$-action 
(\ref{torus action BH}). 
There exist $n$ Young tableaux 
$(\Gamma_1,\Gamma_2,\cdots,\Gamma_n)$ 
by which $(B_a,H_{\dot{A}})$ is labelled. 
Consider an infinitesimal deformation 
$(B_a+\delta B_a,H_{\dot{A}}+\delta H_{\dot{A}})$ 
from this fixed point. 
$(\delta B_a,\delta H_{\dot{A}})$ is  
a tangent vector of the moduli space 
at $(B_a,H_{\dot{A}})$. 
The $U(1)\times U(1)_R$-symmetry rotates 
tangent vectors at the fixed point. 
The $T^2$-action has the form  
\begin{eqnarray}
\delta B_1 && 
\mapsto 
~~e^{i \phi} \gamma(\phi,\theta)^{-1} \delta B_1 
            \gamma(\phi,\theta) , 
\nonumber \\
\delta B_2 && 
\mapsto 
~~e^{i \theta} \gamma(\phi,\theta)^{-1} 
            \delta B_2 \gamma(\phi,\theta) , 
\nonumber \\ 
\delta H_{\dot{1}} && 
\mapsto 
~~e^{i \phi} \gamma(\phi,\theta)^{-1} 
            \delta H_{\dot{1}} , 
\nonumber \\
\delta H_{\dot{2}} && 
\mapsto 
~~e^{-i \theta} \gamma(\phi,\theta)^{-1}
             \delta H_{\dot{2}} ,
\label{torus action tangent}
\end{eqnarray} 
where $\gamma$ is 
the homomorphism from $U(1) \times U(1)_R$ to $U(k)$ 
which is associated with the fixed point 
(cf.(\ref{fixed point})). 
It is possible 
to diagonalize the tangent space with respect to 
$T^2$-action (\ref{torus action tangent}), 
which also provides the diagonalization 
of the hessian of Morse function 
(\ref{Morse function}) at the critical point. 
The eigenspaces with the non-positive 
eigenvalues will generate the maximal cycle 
$C_{(\Gamma_1,\Gamma_2,\cdots,\Gamma_n)}$. 
In the case of 
$(\Gamma_1,\Gamma_2,\cdots,\Gamma_n)$ 
$=(\mbox{$[k]$},\emptyset,\cdots,\emptyset)$  
we obtain (\ref{cycle1}). 
Each one of $a_i$ and $h_{ij}$ in (\ref{cycle1}) 
parametrizes the eigenspace with the non-positive 
eigenvalue in the (holomorphic) tangent space 
at the corresponding fixed point. 
Notice that configuration (\ref{cycle1}) 
itself satisfies $D$-flat condition 
(\ref{ADHM1}) besides stability condition 
(\ref{stability condition}) for any values 
of $a_i$ and $h_{ij}$.

            Taking the viewpoint of D5-branes  
the eigenvalues of $X^i$ (or $B_a$) will describe 
their positions in the four-dimensions. 
Therefore, 
($B_a,H_{\dot{A}}$) in (\ref{cycle1}) 
describes the classical vacuum of $k$ D5-branes 
degenerate at $(z_1,z_2)=(0,0)$.  
It admits the extra 
$nk-1$ complex parameters,  
$a_i$ and $h_{ij}$,  
which constitute the maximal cycle 
$C_{({\scriptsize \mbox{$[k]$}},
\emptyset,\cdots,\emptyset)}$. 
This means that 
{\it any point of the cycle 
$C_{({\scriptsize \mbox{$[k]$}},
\emptyset,\cdots,\emptyset)}$ describes 
the vacuum of $k$ D5-branes degenerate 
at the origin.} 
Notice that, 
though we are describing the vacuum configuration  
$(B_a,H_{\dot{A}})$ using complex symplectic quotient 
(\ref{moduli2}), 
we can always find an element $g$ of $GL(k:{\bf C})$ 
so that 
$(gB_ag^{-1},gH_{\dot{1}},
H_{\dot{2}}^{\dagger}g^{-1})$ 
satisfies $D$-flat condition (\ref{ADHM2}). 
Without loss of generality we can choose 
$g$ such that its lower triangular part are zero 
and therefore $gB_ag^{-1}$ can be regarded as 
upper triangular matrices.  
So, the vacuum 
$(gB_ag^{-1},gH_{\dot{1}},
H_{\dot{2}}^{\dagger}g^{-1})$, 
according to our previous argument, 
goes to zero as $\eta \rightarrow 0$, 
which shows that 
{\it this topological 
cycle disappears at $\eta=0$.}

~

             To study the vacua which belong 
to the maximal cycles it is also useful to describe 
their characteristics in a convenient form. 
For instance, 
$B_a$ in (\ref{cycle1}) satisfy the relations 
\begin{eqnarray}
B_1B_2 &=& B_2B_1,~~~
\nonumber \\ 
B_1^{k}&=&0,~~
\nonumber \\ 
B_2 &=& 
\sum_{i=1}^{k-1}a_i B_{1}^i~~~.
\label{ideal1}
\end{eqnarray}
Similar characterizations 
are also possible for $H_{\dot{A}}$ (\ref{cycle1}). 
In order to describe them    
we first remark that moduli space (\ref{moduli}) 
can be considered as the moduli space of the torsion free 
sheaves of rank $n$ on ${\bf C}P_2$  
\cite{Okonek},\cite{Nakajima}. 
(The complex coordinates $(z_1,z_2)$ of ${\bf C}^2$ 
are identified with the inhomogeneous coordinates 
of ${\bf C}P_2$.) 
The idea is the monad construction of torsion free 
sheaf.  
Consider the following monad complex 
\footnote{${\cal O}_{{\bf C}^2}$ is the structure sheaf of 
${\bf C}^2$.}: 
\begin{eqnarray}
V \otimes {\cal O}_{{\bf C}^2} 
\stackrel{a}{\longrightarrow} 
\begin{array}{c}
V \otimes {\cal O}_{{\bf C}^2} \\ 
\oplus \\ 
V \otimes {\cal O}_{{\bf C}^2} \\ 
\oplus \\ 
W \otimes {\cal O}_{{\bf C}^2} 
\end{array} 
\stackrel{b}{\longrightarrow} 
V \otimes {\cal O}_{{\bf C}^2} 
\label{monad}
\end{eqnarray} 
where 
\begin{eqnarray}
a \equiv 
\left(
\begin{array}{c}
B_1-z_1 \\ 
B_2-z_2 \\ 
\sqrt{2}H_{\dot{2}}^{\dagger} 
\end{array}
\right) 
,~~~~
b \equiv 
\left( 
\begin{array}{ccc}
-B_2+z_2, & B_1-z_1, 
& \sqrt{2}H_{\dot{1}} 
\end{array} 
\right). 
\end{eqnarray}
Notice that it holds $b \cdot a = \mu_{{\bf C}}=0$ 
for any vacuum configuration.  
The corresponding torsion free sheaf 
${\cal E}$ is given by 
\begin{eqnarray}
{\cal E}=Ker~b/ Im~a~~~. 
\label{torsion free sheaf}
\end{eqnarray} 
$GL(k:{\bf C})$-action (\ref{GL(k:C)action}) 
on $(B_a,H_{\dot{A}})$ is noting but 
the linear transform of the bases of $V$. 
The torsion free sheaf ${\cal E}$ given by 
(\ref{torsion free sheaf}) is invariant under this 
transform. Therefore, to each vacuum of $k$ D5-branes 
one can attach a torsion free sheaf 
of rank $n$ uniquely.

                      Let $B_a$ and $H_{\dot{A}}$ 
in complex (\ref{monad}) be those given by equations 
(\ref{cycle1}). 
In such a situation 
an element 
$\left(\begin{array}{c}v_1(z) \\ 
 v_2(z) \\ w(z) \end{array} \right)$ 
of $Ker~b/ Im~a$ satisfies the equation,  
\begin{eqnarray}
\sqrt{2}\sum_{j=1}^n w_j(z)H_{\dot{1}}(e_j)
=(B_2-z_2)v_1(z)-(B_1-z_1)v_2(z), 
\label{ker b}
\end{eqnarray}
where $w(z)=\sum_{j=1}^nw_j(z)e_j$. 
$e_j (1 \leq j \leq n)$ are the bases of 
$W (= {\bf C}^n)$ :   
\begin{eqnarray}
e_1=
\left( \begin{array}{c}
1 \\ 0 \\ \vdots \\ 0 
\end{array}\right),~~
\cdots,~~ 
e_n=
\left( \begin{array}{c}
0 \\ \vdots \\ 0 \\ 1 
\end{array}\right) . 
\end{eqnarray} 
Since it holds that 
\begin{eqnarray}
H_{\dot{1}}(e_j)= 
\sum_{i=1}^k
h_{i j\!-\!1}B_1^{i-1}H_{\dot{1}}(e_1) , 
\end{eqnarray}
for $j \geq 2$, 
we can find out 
$g_j(z) \in {\cal O}_{{\bf C}^2}$ such that 
$H_{\dot{1}}(e_j)=g_j(B)H_{\dot{1}}(e_1)$.  
($~g_1(z) \equiv 1~$).  
Therefore equation (\ref{ker b}) reduces to 
\begin{eqnarray} 
\sqrt{2} 
\sum_{j=1}^nw_j(z)g_j(B)~H_{\dot{1}}(e_1)
=(B_2-z_2)v_1(z)-(B_1-z_1)v_2(z), 
\end{eqnarray}
which, by replacing $z_a$ with $B_a$, 
implies  
\begin{eqnarray}
\sum_{j=1}^nw_j(B)g_j(B)H_{\dot{1}}(e_1)=0. 
\end{eqnarray} 
Because $H_{\dot{1}}(e_1)$ is cyclic 
with respect to $B_a$ and 
$B_a$ themselves are commutative,  
we can conclude that  
\begin{eqnarray}
\sum_{j=1}^nw_j(B)g_j(B)=0 .
\label{relation1}
\end{eqnarray}   
Equation (\ref{relation1}) 
may be simplified by introducing the analogue 
of gauge transformation 
\begin{eqnarray} 
w=
\left( \begin{array}{c}
w_1 \\
\vdots \\ 
w_n 
\end{array} \right) 
\mapsto 
\tilde{w}=
\left( \begin{array}{c}
\tilde{w}_1 \\
\vdots \\ 
\tilde{w}_n 
\end{array} \right)
=G w~~~,
\end{eqnarray} 
where the ``gauge transform" $G(z)$ has the form 
\begin{eqnarray}
G(z) =
\left( 
\begin{array}{ccccc}
1      & g_2(z) & g_3(z) & \ldots & g_n(z) \\
0      & 1      & 0      & \ldots & 0      \\ 
\vdots & \ddots & \ddots & \ddots & \vdots \\ 
\vdots &        & \ddots & \ddots & 0      \\ 
0      & \ldots & \ldots & 0      & 1
\end{array}\right)~~.
\label{gauge}
\end{eqnarray}
Notice that each $g_j(z) \in {\cal O}_{{\bf C}^2}$ 
is given by the relation, 
$H_{\dot{1}}(e_j)=g_j(B)H_{\dot{1}}(e_1)$. 
The data of $H_{\dot{A}}$ are now encoded 
into the ``gauge transform" $G$. 
By this transformation, 
equation (\ref{relation1}) becomes 
equivalent to the following constraint  
on the first component of $\tilde{w}$, 
\begin{eqnarray}
\tilde{w}_1(B)=0. 
\end{eqnarray}
Due to relations (\ref{ideal1}) 
any polynomial which satisfies this constraint 
can be given by the combinations  
\begin{eqnarray}
r(z)z_1^{k}+s(z)(z_2-\sum_{i=1}^{k-1}a_iz_1^{i}), 
\end{eqnarray} 
where $r(z),s(z) \in {\cal O}_{{\bf C}^2}$. 
Therefore it seems very plausible that 
one can distinguish 
$B_a$ and $H_{\dot{A}}$ (\ref{cycle1}) 
by relations (\ref{ideal1}) and 
``gauge transform" (\ref{gauge}).

~

          $(B_a,H_{\dot{A}})$ given in 
(\ref{cycle1}) describes the vacuum of $k$ D5-branes 
degenerate at the origin $(0,0)$. 
These vacua form the maximal cycle 
$C_{({\scriptsize \mbox{$[k]$}},
\emptyset,\cdots,\emptyset)}$.  
By changing the eigenvalues of $B_a$ from $0$ to 
$z_a$ in (\ref{cycle1}) 
it provides the vacuum configuration that 
$k$ D5-branes are overlapping at $P=(z_1,z_2)$. 
The additional parameters in (\ref{cycle1}) also 
constitute a cycle of the moduli space but 
it can be topologically identified with 
$C_{({\scriptsize \mbox{$[k]$}},\emptyset,
\cdots,\emptyset)}$. 
Let us consider the vacua of 
$k=k_1+\cdots+k_l$ D5-branes  
($k_1 \geq k_2 \geq \cdots \geq k_l$) 
in which each $k_i$ pieces are degenerate 
at $P_i=(z_1^{(i)},z_2^{(i)})$. 
Among them we shall concentrate on the configuration 
obtained by superposing the vacua of 
$k_i$ D5-branes ($1 \leq i \leq l$) 
having forms (\ref{cycle1}) 
with their eigenvalues shifted. 
Explicitly such a configuration may be written as   
\begin{eqnarray}
B_a=
\left( \begin{array}{ccc}
B_a^{(1)} &        &           \\
          & \ddots &           \\   
          &        & B_a^{(l)}  
\end{array} \right)
~~~,~~~ 
H_{\dot{A}}= 
\left( \begin{array}{c} 
H_{\dot{A}}^{(1)} \\ 
\vdots \\ 
H_{\dot{A}}^{(l)} 
\end{array} \right)~~. 
\label{cycle2}
\end{eqnarray} 
$B_a^{(i)}$ and $H_{\dot{A}}^{(i)}$ are respectively 
$k_i \times k_i$ and $k_i \times n$ complex matrices 
such that $(B_a^{(i)},H_{\dot{A}}^{(i)})$ 
describes the vacuum of $k_i$ D5-branes 
overlapping at $P_i=(z_1^{(i)},z_2^{(i)})$ 
which is given (cf. (\ref{cycle1})) by  
\begin{eqnarray}
&&
B_1^{(i)} = 
\left( 
\begin{array}{ccccc}
z_1^{(i)} & 1         & 0       & \ldots & 0        \\
0         & z_1^{(i)} & 1       & \ddots & \vdots   \\ 
\vdots    &  0        & \ddots  & \ddots & 0        \\
\vdots    & \vdots    & \ddots  & \ddots & 1        \\ 
0         &  0        & \ldots  &  0     & z_1^{(i)}
\end{array} 
\right)
~~,~~  
B_2^{(i)} = 
\left( 
\begin{array}{ccccc}
z_2^{(i)}&a_1^{(i)}& a_2^{(i)}&\ldots& a_{k_i\!-\!1}^{(i)}\\
0       &z_2^{(i)}&a_1^{(i)}&\ddots& \vdots \\ 
\vdots  &  0     & \ddots  & \ddots  & a_2^{(i)}  \\
\vdots  & \vdots & \ddots  & \ddots  & a_1^{(i)}  \\ 
0       &  0     & \ldots  &  0      & z_2^{(i)}
\end{array}
\right) 
\nonumber 
\\ 
&&
H_{\dot{1}}^{(i)} =
\left( 
\begin{array}{cccc}
0      & h_{11}^{(i)}& \ldots & h_{1 n\!-\!1}^{(i)} \\
\vdots & \vdots    &        & \vdots \\ 
\vdots & \vdots    &        & \vdots \\
0 &h_{k_i\!-\!1 1}^{(i)}&\ldots &h_{k_i\!-\!1 n\!-1}^{(i)} \\ 
1      & h_{k_i 1}^{(i)} & \ldots & h_{k_i n\!-\!1}^{(i)}
\end{array} 
\right) 
~~, ~~
H_{\dot{2}}^{(i)}=0~~. 
\label{element of cycle2}
\end{eqnarray}
For each $i$, vacuum configuration (\ref{element of cycle2}) 
admits the additional $nk_i-1$ complex parameters,  
$a^{(i)}$ and $h^{(i)}$. 
They form the maximal cycle 
$C_{({\scriptsize \mbox{$[k_i]$}},\emptyset,\cdots,\emptyset)}$  
in the moduli space ${\cal M}(k_i)$. 
One can expect that these parameters, 
considering them as the additional parameters for the 
vacuum of $k=k_1+\cdots+k_i$ D5-branes,  
also constitute an appropriate cycle in the moduli space 
${\cal M}(k)$.

              It is important to note that explicit form  
(\ref{cycle2}) does not always describe a vacuum 
when $P_i$ coincides with $P_j$ for some $i$ and $j$. 
This is because 
the first column vector of $H_{\dot{1}}$ in  
(\ref{cycle2}) is no longer a cyclic vector 
when $(z_1^{(i)},z_2^{(i)})=(z_1^{(j)},z_2^{(j)})$. 
Notice that, as far as all the $l$-points 
$P_1,\cdots,P_l$ are different from one another, 
the first column vector of $H_{\dot{1}}$ 
is cyclic and therefore configuration 
(\ref{cycle2}) describes the vacuum 
of $k$ D5-branes for any values of 
the additional complex parameters. 
In the case when some of the $l$-points coincide, 
to guarantee stability (\ref{stability condition}) of 
configuration (\ref{cycle2}),  
it may become necessary to introduce a subspace of 
$Im~H_{\dot{1}}$ spanned by not only the first column 
vector but also the other column vectors of $H_{\dot{1}}$ 
and then to consider the possibility whether 
the vectors of this subspace besides the vectors obtained 
from them by the successive actions of $B_a$ can span $V$. 
Though this prescription  for restoring the stability  
of the configuration seems to work, the introduction 
of the column vectors of $H_{\dot{1}}$ other than 
the first one turns out to freeze some 
of the additional complex parameters in (\ref{cycle2}). 
(In fact some of these parameters are absorbed into the 
$GL(k:{\bf C})$-symmetry.)

            There is another prescription to 
restore the stability condition. 
Take an element $q$ of $GL(k:{\bf C})$ 
which depends on the $l$-points 
$P_1,\cdots,P_l$ and becomes singular 
when some of these points coincide. 
Transform (\ref{GL(k:C)action}) 
of vacuum (\ref{cycle2}) by $q$ is gauge-equivalent to 
the original configuration 
as far as these $l$-points are different 
from one another. 
It might be possible to choose $q$ of 
$GL(k:{\bf C})$ so that the transform by $q$ 
is still well-defined without missing any complex 
parameters in (\ref{cycle2}) even when some of 
the $l$-points coincide. 
The complex parameters will be rescaled such that 
the singularities of $q$ are absorbed into them. 
One may say that such an element $q$ of 
$GL(k:{\bf C})$ gives the change of coordinates 
of the cycle.

            These modified forms (which are obtained 
by appropriate singular gauge transformations) 
will appear naturally in our approach. 
It is because our description of  
the topological cycles of the moduli space 
is based on the consideration of  
the fixed points of the $T^2$-action. 
Since the $U(1)\times U(1)_R$-symmetry rotates 
$P_i=(z_1^{(i)},z_2^{(i)})$ in the way given by 
(\ref{torus action}), 
the cycle formed by the complex parameters in (\ref{cycle2}) 
will be captured in our treatment 
by examining the limit that the $l$-points in 
(\ref{cycle2}) approach to one another 
and go to the origin $(0,0)$. 
In fact, we will show that 
{\it the vacuum which belongs to the maximal 
cycle $C_{(\Gamma,\emptyset,\cdots,\emptyset)}$ 
with $\Gamma= \mbox{$[k_1,\cdots,k_l]$}$  
can be identified with  
(\ref{cycle2}) under the limit 
that all the $l$-different points 
$P_1,\cdots, P_l$ in (\ref{cycle2})
go to the origin $(0,0)$. }  
For this identification 
it is necessary to know 
how the vacuum configuration in which   
each $k_i$ D5-branes are degenerate at $P_i$  
behaves when all these $l$-points approach to 
the origin and 
then to compare it with the configuration 
which belongs to the maximal cycle 
mentioned above.

~

              It may be enough to consider the vacuum 
in which $m$ D5-branes are overlapping at a point $P$ 
in the four-dimensions while the other D5-branes 
are degenerate at another point, say $Q$, 
and then to compare this configuration, 
when $P$ goes to $Q$,  
with the vacuum 
which belongs to the corresponding maximal cycle.  
For definiteness 
let us suppose that $m$ and $m+m'$ D5-branes 
are degenerate respectively at 
$P=(0,\lambda)$ and $Q=(0,0)$. 
$\lambda \neq 0$. 
Their vacuum is denoted   
by $(\hat{B}_a,\hat{H}_{\dot{A}})$. 
$\hat{B}_a$ have the block-diagonal form 
$\left( \begin{array}{cc}
\hat{B}_a^{(1)} & 0 \\
0  & \hat{B}_a^{(2)}
\end{array}\right)$ 
where $\hat{B}_a^{(1)}$ and $\hat{B}_a^{(2)}$ 
are respectively $m \times m$ and $(m+m')\times(m+m')$ 
complex matrices. 
$\hat{H}_{\dot{A}}$ also have the form 
$\left(\begin{array}{c}
\hat{H}_{\dot{A}}^{(1)} \\
\hat{H}_{\dot{A}}^{(2)} 
\end{array}\right)$ 
where 
$\hat{H}_{\dot{A}}^{(1)}$ 
and 
$\hat{H}_{\dot{A}}^{(2)}$ 
are respectively 
$m \times n$ and $(m+m')\times n$ 
complex matrices. 
The constituent $(\hat{B}_a^{(1)},\hat{H}_{\dot{A}}^{(1)})$ 
($~(\hat{B}_a^{(2)},\hat{H}_{\dot{A}}^{(2)})~$) 
describes the contribution of $m$ ($~m+m'~$) 
D5-branes degenerate at $P$ ($~Q~$).  
The explicit form is given in Table 1.  
$(\hat{B}_a,\hat{H}_{\dot{A}})$ in Table 1
admits $n(2m+m')-2$ complex parameters 
$\hat{a}_j,\hat{b}_j,\hat{d}_j$ and $\hat{h}_{ij}$.  
Among them, 
$\hat{b}_j$ and $\hat{h}_{ij}$ $(1 \leq i \leq m)$ 
are the additional parameters  associated with 
the configuration of the first $m$ D5-branes  
while $\hat{a}_j$, $\hat{d}_j$ and  
$\hat{h}_{ij}$ $(m+1 \leq i \leq 2m+m')$ 
are those associated with the $m+m'$ 
D5-branes at $Q$.

            For this vacuum one can provide a 
characterization similar to (\ref{ideal1}). 
From the form given in Table 1 the following 
relations among $\hat{B}_a$ can be found out : 
\begin{eqnarray} 
\hat{B}_1^{m+m'}&=&0,  
\nonumber 
\\ 
\hat{B}_1^m 
\left( 
\hat{B}_2-\sum_{j=1}^{m'-1}\hat{a}_j\hat{B}_1^j 
\right)&=&0,~~
\nonumber 
\\
\left(
\hat{B}_2-\sum_{j=1}^{m-1}\hat{b}_j\hat{B}_1^j-\lambda 
\right) 
\left(
\hat{B}_2-\sum_{j=1}^{m'-1}\hat{a}_j\hat{B}_1^j \right) 
&=&
\sum_{j=1}^m\hat{c}_j\hat{B}_1^{m'+j-1}, 
\nonumber \\
\hat{B}_1\hat{B}_2 
&=& 
\hat{B}_2\hat{B}_1 
\label{ideal2}
\end{eqnarray}
where $\hat{c}_j$ ($1 \leq j \leq m$) are 
introduced by 
\begin{eqnarray}
\hat{c}_j \equiv 
-\lambda \hat{d}_j+\sum_{r+s=j}
(\hat{a}_r-\hat{b}_r)\hat{d}_s.   
\label{parameter hatc}
\end{eqnarray} 
Since $\lambda$ is not equal to zero,   
these $\hat{c}_j$ are equivalent to the parameters 
$\hat{d}_j$.

            Let us examine the limit that 
$P=(0,\lambda)$ goes to $Q=(0,0)$. 
The configuration obtained from 
$(\hat{B}_a,\hat{H}_{\dot{A}})$ 
in Table 1 by setting $\lambda$ equal to zero   
does not always describe a vacuum of D5-branes.   
This is because, when $\lambda =0$,  
the first column vector of $\hat{H}_{\dot{1}}$ 
in Table 1 is not cyclic with respect to $\hat{B}_a$ 
and it becomes necessary to compensate it 
with the other column vectors of $\hat{H}_{\dot{1}}$ 
in order to satisfy stability condition 
(\ref{stability condition}). 
With this compensation some of $\hat{h}_{ij}$ 
in $\hat{H}_{\dot{1}}$ are frozen or absorbed into the 
$GL(2m+m':{\bf C})$-symmetry. 
The number of the additional complex parameters 
will decrease from $(2m+m')n-2$. 
To avoid such a decrease of parameters 
we need to change our perspective. 
We begin by describing the maximal cycle 
$C_{({\scriptsize \mbox{$[m+m',m]$}}, 
\emptyset,\cdots,\emptyset)}$.  
The dimensions of this cycle  
can be read as follows 
\begin{eqnarray} 
dim~ C_{({\scriptsize \mbox{$[m+m',m]$}},
\emptyset,\cdots,\emptyset)} 
&=& 
2(n-1)+
2 \left\{ 
n(2m+m'-2)+(n-1) 
\right\}  
\nonumber \\ 
&=& 
2n(2m+m')-4.
\end{eqnarray} 
The vacuum configuration 
$(B_a,H_{\dot{A}})$ in  
$C_{({\scriptsize \mbox{$[m+m',m]$}},
\emptyset,\cdots,\emptyset)}$ 
is given in Table 2. 
It has $n(2m+m')-2$ complex parameters 
$a_j$, $b_j$, $c_j$ and $h_{ij}$.  
These parameters constitute the maximal cycle 
$C_{({\scriptsize \mbox{$[m+m',m]$}},
\emptyset,\cdots,\emptyset)}$. 
After some manipulation we can find out 
$B_a$ in Table 2 satisfy the relations 
\begin{eqnarray}
B_1^{m+m'}&=&0,  
\nonumber 
\\ 
B_1^m 
\left( 
B_2-\sum_{j=1}^{m'-1}a_jB_1^j 
\right)&=&0,~~
\nonumber 
\\
\left(
B_2-\sum_{j=1}^{m-1}b_jB_1^j  
\right) 
\left(
B_2-\sum_{j=1}^{m'-1}a_jB_1^j \right) 
&=&
\sum_{j=1}^mc_jB_1^{m'+j-1} ,
\nonumber \\ 
B_1B_2 
&=& 
B_2B_1 . 
\label{ideal3}
\end{eqnarray}

               Now we can investigate 
the behavior of the vacuum 
$(\hat{B}_a,\hat{H}_{\dot{A}})$ 
when $P$ goes to $Q$. 
As we have mentioned,  
we should not set $\lambda=0$ 
for the explicit form in Table 1. 
But relations (\ref{ideal2}) themselves 
might be handled under this limit. 
We first rescale $\hat{c}_j$ (\ref{parameter hatc}) 
such that they do not depend on $\lambda$ and 
then introduce the new parameters in stead of 
$\hat{a}_j,\hat{b}_j$ and $\hat{d}_j$ : 
\begin{eqnarray}
a_j &=& \hat{a}_j, 
\nonumber \\ 
b_j &=& \hat{b}_j, 
\nonumber \\ 
c_j &=& 
\hat{c}_j(\hat{a},\hat{b},\hat{d}:\lambda). 
\label{newparameter}
\end{eqnarray} 
With fixing these new parameters,  
relations (\ref{ideal2}) become (\ref{ideal3}) 
as $\lambda$ goes to zero ! 
Notice that equations (\ref{newparameter}) 
can be regarded as the change of coordinates 
as far as $\lambda \neq 0$. 
These new parameters themselves will survive 
even at $\lambda =0$. 
The payoff for this transmutation of the parameters 
is the change of realization of the vacuum 
from those given in Table 1 to those given in Table 2.  
To complete the discussion one should also introduce  
similar rescalings of the parameters $\hat{h}_{ij}$. 
Though the investigation of 
``gauge transform" (cf. (\ref{gauge})) provides 
a nice description of the rescalings 
for these parameters, 
it becomes complicated and hard to give a general theory. 
So we convince the reader by giving some examples.

\subsubsection*{Example 1 : $(m,m')=(1,0)$} 

                    Let us consider the case of 
two D5-branes located at two different points, 
say $P$ and $Q$. For definiteness we set 
$P=(0,\lambda)$ and $Q=(0,0)$. 
$\lambda \neq 0$. 
The corresponding vacuum configuration, 
which is denoted by 
$(\hat{B}_a,\hat{H}_{\dot{A}})$, 
has the form 
\begin{eqnarray} 
&&
\hat{B}_1 =  
\left(
\begin{array}{cc} 
0 & 0 \\ 
0 & 0 
\end{array} 
\right) 
~~,~~ 
\hat{B}_2= 
\left( 
\begin{array}{cc} 
\lambda & 0 \\ 
0 & 0 
\end{array}
\right) ,
\nonumber \\
&&\hat{H}_{\dot{1}} =
\left( 
\begin{array}{cccc}
1 & \hat{h}_{11} & \cdots & \hat{h}_{1 n\!-\!1} 
\\ 
1 & \hat{h}_{21} & \cdots & \hat{h}_{2 n\!-\!1} 
\end{array}
\right)~~,~~ 
\hat{H}_{\dot{2}}=0.  
\label{example1-1}
\end{eqnarray} 
$\hat{h}_{1j}$ and $\hat{h}_{2j}$ are 
respectively the additional parameters 
associated with the D5-brane located at 
$P$ and $Q$. 
Notice that $\hat{B}_2$ in 
(\ref{example1-1}) satisfies the relation 
\begin{eqnarray}
\hat{B}_2(\hat{B}_2-\lambda)=0 .
\label{eq of B example1-1} 
\end{eqnarray}
As far as $\lambda \neq 0$,   
$\hat{H}_{\dot{1}}(e_1)
=\left( \begin{array}{c} 
1 \\ 1 
\end{array}
\right)$ is cyclic with respect to $\hat{B}_2$. 
In particular  
$\hat{H}_{\dot{1}}(e_j)$ $(j \geq 2)$ 
are obtainable from 
$\hat{H}_{\dot{1}}(e_1)$ 
by appropriate actions of 
$\hat{B}_2$ :  
\begin{eqnarray}
\hat{H}_{\dot{1}}(e_j)=
\left\{ 
\frac{\hat{h}_{1j\!-\!1}-\hat{h}_{2 j\!-\!1}}{\lambda} 
\hat{B}_2+\hat{h}_{2j\!-\!1} 
\right\}
\hat{H}_{\dot{1}}(e_1). 
\end{eqnarray}
With this expression of $\hat{H}_{\dot{1}}(e_j)$ 
one can see that the corresponding 
``gauge transform" (cf. (\ref{gauge}) ), 
which we write as $\hat{G}$,  
is given by 
\begin{eqnarray}
\hat{G}(z) = 
\left( 
\begin{array}{ccccc}
1      & \hat{g}_2(z) & \hat{g}_3(z) & \ldots & 
\hat{g}_n(z) \\
0      & 1      & 0      & \ldots & 0      \\ 
\vdots & \ddots & \ddots & \ddots & \vdots \\ 
\vdots &        & \ddots & \ddots & 0      \\ 
0      & \ldots & \ldots & 0      & 1
\end{array}\right),  
\label{gauge example1-1}
\end{eqnarray}
where $\hat{g}_j(z)$ 
$(j \geq 2)$ are 
\begin{eqnarray}
\hat{g}_j(z)=
\frac{\hat{h}_{1j\!-\!1}-\hat{h}_{2 j\!-\!1}}{\lambda} 
z_2+\hat{h}_{2j\!-\!1}. 
\label{hatg example1-1}
\end{eqnarray}

                 At this stage let us remark 
on the configuration obtained from (\ref{example1-1}) 
by setting $\lambda =0$.  
The first column vector of $\hat{H}_{\dot{1}}$, 
$\left(\begin{array}{c}
1 \\ 
1 
\end{array}\right)$, is no longer cyclic. 
In order that this configuration describes a vacuum 
one must impose an appropriate constraint 
on the parameters $\hat{h}_{ij}$ 
so that it satisfies stability condition 
(\ref{stability condition}). 
It is sufficient if 
$\left(\begin{array}{c}
\hat{h}_{11} \\
\hat{h}_{21} 
\end{array}\right)$ 
besides 
$\left(\begin{array}{c}
1 \\ 
1 \\
\end{array}\right)$ 
span ${\bf C}^2 (\equiv V)$. 
In such a situation one can take these two vectors as 
the bases of $V$. In terms of these new bases, 
vacuum (\ref{example1-1}), setting $\lambda=0$,  
acquires the form, $\hat{B}_a=\hat{H}_{\dot{2}}=0$ and 
$\hat{H}_{\dot{1}}=
\left(\begin{array}{ccc}
0 & 1 & * \\ 
1 & 0 & * 
\end{array}\right)$,  
which means the parameters 
$\hat{h}_{11,21}$ are absorbed into the 
$GL(2:{\bf C})$-symmetry.

                   Nextly we describe the vacuum   
which belongs to the maximal cycle 
$C_{({\scriptsize \mbox{$[1,1]$}}, 
\emptyset,\cdots,\emptyset)}$. 
We shall write it $(B_a,H_{\dot{A}})$. 
It can be realized by  
\begin{eqnarray}
&& 
B_1=0~~,~~ 
B_2=
\left( 
\begin{array}{cc}
0 & 1 \\
0 & 0
\end{array}\right), 
\nonumber \\ 
&&
H_{\dot{1}} = 
\left( \begin{array}{cccc}
0 & h_{11} & \cdots & h_{1 n\!-\!1} \\ 
1 & h_{21} & \cdots & h_{2 n\!-\!1} 
\end{array} \right) ~~,~~ 
H_{\dot{2}}=0, 
\label{example1-2}
\end{eqnarray}
where $2(n-1)$ complex parameters 
$h_{ij}$ constitue the maximal cycle. 
Notice that $B_2$ satisfy the relation,  
\begin{eqnarray}
B_2^2=0~~.
\label{eq of B example1-2}
\end{eqnarray} 
As for ``gauge transform" $G$ in 
(\ref{gauge}), 
the elements of $G$ turn out to be given by 
\begin{eqnarray}
g_j(z)=
h_{1j\!-\!1}z_2+h_{2j\!-\!1} .
\label{g example1-2}
\end{eqnarray}

               Let us study the behavior of 
configuration (\ref{example1-1}) 
when $P$ goes to $Q$ 
by considering rescalings of    
the complex parameters in (\ref{example1-1}).  
Namely we rescale $\hat{h}_{ij}$ such that 
$\hat{h}_{2j}$ as well as 
$(\hat{h}_{1j}-\hat{h}_{2j})/\lambda$ 
do not depend on $\lambda$.   
We may introduce the new parameters 
in stead of $\hat{h}_{ij}$ :   
\begin{eqnarray}
h_{1j}&=&
\frac{\hat{h}_{1j}-\hat{h}_{2j}}{\lambda}~~, 
\nonumber \\
h_{2j}&=&
\hat{h}_{2j}.  
\label{rescale example1}
\end{eqnarray}
Equations (\ref{rescale example1}) 
can be also 
regarded as the change of coordinates 
as far as $\lambda \neq 0$. 
Taking the element $\hat{q}(\lambda)$ of $GL(2:{\bf C})$,  
\begin{eqnarray}
\hat{q}(\lambda)=
\left( \begin{array}{cc}
1/\lambda & -1/\lambda \\ 
0 & 1 
\end{array}\right)~~  
\end{eqnarray}
the transform of (\ref{example1-1}) 
by $\hat{q}(\lambda)$
becomes 
\begin{eqnarray} 
\hat{q}(\lambda)\hat{B}_1\hat{q}(\lambda)^{-1}
&=& 
0~,~~ 
\hat{q}(\lambda)\hat{B}_2\hat{q}(\lambda)^{-1}~=~
\left(\begin{array}{cc}
\lambda & 1 \\ 
0       & 0 
\end{array}\right)~~, 
\nonumber \\ 
\hat{q}(\lambda)H_{\dot{1}}
&=& 
\left(\begin{array}{cccc} 
0 & 
\frac{\hat{h}_{11}\!-\!\hat{h}_{21}}{\lambda} & 
\cdots & 
\frac{\hat{h}_{1 n\!-\!1}\!-\!\hat{h}_{2 n\!-\!1}}
{\lambda} \\ 
1 & 
\hat{h}_{21} & 
\cdots & \hat{h}_{2 n\!-\!1} 
\end{array} \right). 
\label{example1-3}
\end{eqnarray}
Rewritting  configuration (\ref{example1-3}) 
by rescaled parameters (\ref{rescale example1}), 
we can find that it 
coincides with (\ref{example1-2}) 
as $\lambda$ goes to zero. 
This shows that the rescaled parameters 
survive even at $\lambda=0$ and that,  
due to the singularity at $\lambda=0$ in 
$\hat{q}(\lambda)$,  
one should change the realization from 
(\ref{example1-1}) to (\ref{example1-2}), 
at least, at $P=Q$.

\subsubsection*{Example 2 : $(m,m')=(1,1)$} 

        As the second example we shall consider 
the case of three D5-branes one of which is 
located at $P=(0,\lambda)$ and 
the other two are overlapping at $Q=(0,0)$. 
$\lambda \neq 0$. 
The corresponding vacuum 
$(\hat{B}_a,\hat{H}_{\dot{A}})$ has the form 
\begin{eqnarray}
&&
\hat{B}_1=
\left( 
\begin{array}{ccc}
0 & 0 & 0 \\
0 & 0 & 1 \\
0 & 0 & 0 
\end{array}
\right) ~~~,~~~ 
\hat{B}_2 
=
\left( 
\begin{array}{ccc}
\lambda & 0 & 0 \\
0 & 0 & \hat{a} \\
0 & 0 & 0 
\end{array}
\right)~~~, 
\nonumber \\
&&
\hat{H}_{\dot{1}} =
\left( 
\begin{array}{cccc}
1 & \hat{h}_{11} & \cdots & \hat{h}_{1 n\!-\!1} 
\\ 
0 & \hat{h}_{21} & \cdots & \hat{h}_{2 n\!-\!1} 
\\
1 & \hat{h}_{31} & \cdots & \hat{h}_{3 n\!-\!1} 
\end{array}
\right)~~,~~~ 
\hat{H}_{\dot{2}}=0~~~,
\label{example2-1}
\end{eqnarray} 
where $\hat{a}$ and $\hat{h}_{ij}$ are 
$3n-2$ complex parameters. 
Taking the diagonal elements of $\hat{B}_a$ 
into account one can say $\hat{h}_{1j}$  
are the additional parameters 
associated with the first D5-brane 
located at $P$ while  
$\hat{a}$ besides $\hat{h}_{2j}$ and $\hat{h}_{3j}$ 
are those for the two D5-branes overlapping at $Q$. 
It is easy to check that 
$\hat{B}_a$ in (\ref{example2-1}) satisfy the 
relations 
\begin{eqnarray}
&& 
\hat{B}_1^2=0 ~~, 
\nonumber \\
&& 
(\hat{B}_2-\lambda)(\hat{B}_2-\hat{a}\hat{B}_1)=0~~,
\nonumber \\ 
&& 
\hat{B}_1\hat{B}_2=\hat{B}_2\hat{B}_1=0~~.
\label{eq of B example2-1}
\end{eqnarray}
Notice that  
$\hat{H}_{\dot{1}}(e_1)$ is cyclic with respect to 
$\hat{B}_a$.  
In particular one can write down 
$\hat{H}_{\dot{1}}(e_j)$ $(j \geq 2)$ in the form 
\begin{eqnarray}
&&
\hat{H}_{\dot{1}}(e_j)=
\left\{
\frac{\hat{h}_{1j\!-\!1}-\hat{h}_{3j\!-\!1}}{\lambda}
\hat{B}_2+ 
\left( 
\hat{h}_{2j\!-\!1}-
\frac{\hat{a}(\hat{h}_{1j\!-\!1}-\hat{h}_{3j\!-\!1})}
{\lambda}
\right)
\hat{B}_1+\hat{h}_{3j\!-\!1}
\right\} 
\hat{H}_{\dot{1}}(e_1)~~.
\nonumber \\
&&~~ 
\label{Hj example2-1}
\end{eqnarray}
The ``gauge transform" $\hat{G}$, 
which has a form similar to (\ref{gauge example1-1}), 
is now given by the polynomials 
$\hat{g}_j(z)$ $(j \geq 2)$ 
determined from (\ref{Hj example2-1}):  
\begin{eqnarray}
\hat{g}_j(z)=
\frac{\hat{h}_{1j\!-\!1}-\hat{h}_{3j\!-\!1}}{\lambda}
z_2+ 
\left( 
\hat{h}_{2j\!-\!1}-
\frac{\hat{a}(\hat{h}_{1j\!-\!1}-\hat{h}_{3j\!-\!1})}
{\lambda}
\right)
z_1+\hat{h}_{3j\!-\!1} ~~~. 
\label{hatg example2-1}
\end{eqnarray}

      Let us remark on the configuration 
obtained from (\ref{example2-1}) 
by setting $\lambda=0$.  
In order that this configuration describes a vacuum 
it is sufficient that the second column vector of 
$\hat{H}_{\dot{1}}$ besides 
$\left(\begin{array}{c}
1 \\ 
0 \\ 
1 
\end{array}\right)$ and 
$\hat{B}_1
\left(\begin{array}{c}
1 \\ 
0 \\ 
1 
\end{array}\right)$ 
span ${\bf C}^3 (\equiv V)$. 
This condition gives the constraint, 
$\hat{h}_{11}\neq \hat{h}_{31}$. 
If it is satisfied, 
one can take a new bases of $V$ so that 
vacuum (\ref{example2-1}) (with setting $\lambda=0$) 
acquires the form 
\begin{eqnarray}
&&
\hat{B}_1=
\left( 
\begin{array}{ccc}
0 & 1 & \hat{h}_{31} \\
0 & 0 & 0 \\
0 & 0 & 0 
\end{array}
\right) ~~~,~~~ 
\hat{B}_2 
=
\left( 
\begin{array}{ccc}
0& \hat{a} & \hat{a}\hat{h}_{31} \\
0 & 0 & 0 \\
0 & 0 & 0 
\end{array}
\right)~~~, 
\nonumber \\
&&
\hat{H}_{\dot{1}} =
\left( 
\begin{array}{cccc}
0 & 0 & * & * 
\\ 
0 & 1 & * & * 
\\
1 & 0 & * & * 
\end{array}
\right)~~,~~~ 
\hat{H}_{\dot{2}}=0~~~,
\end{eqnarray}
which means that the parameters 
$\hat{h}_{11,21}$ are absorbed into 
the $GL(3:{\bf C})$-symmetry.

             Nextly we describe the vacuum 
$(B_a,H_{\dot{A}})$ which belongs 
to the maximal cycle 
$C_{({\scriptsize \mbox{$[$}2,1\mbox{]}},
\emptyset,\cdots,\emptyset)}$. 
It can be realized by the form 
\begin{eqnarray}
&&
B_1  = 
\left( 
\begin{array}{ccc}
0 & 0 & 0 \\
0 & 0 & 1 \\
0 & 0 & 0 
\end{array}
\right) ~~~,~~~ 
B_2 
=
\left( 
\begin{array}{ccc}
0 & 0 & 1 \\
a & 0 & 0 \\
0 & 0 & 0 
\end{array}
\right)~~~, 
\nonumber \\
&&
H_{\dot{1}} =
\left( 
\begin{array}{cccc}
0 & h_{11} & \cdots & h_{1 n\!-\!1} 
\\ 
0 & h_{21} & \cdots & h_{2 n\!-\!1} 
\\
1 & h_{31} & \cdots & h_{3 n\!-\!1} 
\end{array}
\right)~~,~~~ 
H_{\dot{2}}=0~~~,
\label{example2-2}
\end{eqnarray} 
where $3n-2$ complex parameters 
$a$ and $h_{ij}$ constitue the maximal cycle 
$C_{({\scriptsize \mbox{$[2,1]$}},
\emptyset,\cdots,\emptyset)}$.  
One can check $B_a$ satisfy the relations 
\begin{eqnarray}
&& 
B_1^2=0~~,
\nonumber \\
&& 
B_2^2-aB_1=0~~,
\nonumber \\
&& 
B_1B_2=B_2B_1=0~~.
\label{eq of B example2-2}
\end{eqnarray}
As for ``gauge transform" $G$ (\ref{gauge}), 
its elements turn out to be given by 
\begin{eqnarray}
g_j(z)=h_{1j\!-\!1}z_2+h_{2j\!-\!1}z_1+
h_{3j\!-\!1}. 
\label{g example2-2}
\end{eqnarray}

              The behavior of configuration 
(\ref{example2-1}) when $P$ goes to $Q$ will be studied 
taking account of the rescaling of parameters. 
We first rescale $\hat{a}$ such that 
$\lambda \hat{a}$ does not depend on $\lambda$.  
We may introduce the new parameter by 
\begin{eqnarray}
a = -\lambda \hat{a}. 
\label{rescale1 example2}
\end{eqnarray} 
Notice that relations (\ref{eq of B example2-1}), 
with fixing $a$,  
become (\ref{eq of B example2-2}) 
as $\lambda$ goes to $0$. 
One can also rescale $\hat{h}_{ij}$ such that 
\begin{eqnarray}
&&
h_{1j} = 
\frac{\hat{h}_{1j}-\hat{h}_{3j}}{\lambda}~~, 
\nonumber \\ 
&& 
h_{2j} = 
\hat{h}_{2j}+
\frac{a(\hat{h}_{1j}-\hat{h}_{3j})}{\lambda^2}~~, 
\nonumber \\
&& 
h_{3j} = 
\hat{h}_{3j}~~
\label{rescale2 example2}
\end{eqnarray}
do not depend on $a$ and $\lambda$. 
Elements (\ref{hatg example2-1}) 
in the gauge transform $\hat{G}$, 
expressing them in terms of these rescaled $h_{ij}$,  
coincide with (\ref{g example2-2}). 
Equations 
(\ref{rescale1 example2}) 
and (\ref{rescale2 example2}) 
can be also regarded as the change of coordinates 
as far as $\lambda \neq 0$. 
Introduce the element $\hat{q}(\lambda)$ of 
$GL(3:{\bf C})$ by 
\begin{eqnarray}
\hat{q}(\lambda)=
\left(\begin{array}{ccc}
1/\lambda & 0 & -1/\lambda \\ 
-\hat{a}/\lambda & 1 & \hat{a}/\lambda \\
0 & 0 & 1 
\end{array}\right)~~. 
\end{eqnarray}
The transform of (\ref{example2-1}) by 
$\hat{q}(\lambda)$ becomes  
\begin{eqnarray}
\hat{q}(\lambda) \hat{B}_1 \hat{q}(\lambda)^{-1}
&=& 
\left( \begin{array}{ccc}
0 & 0 & 0 \\
0 & 0 & 1 \\
0 & 0 & 0 
\end{array}
\right) ~,~  
\hat{q}(\lambda) \hat{B}_2 \hat{q}(\lambda)^{-1}
~=~ 
\left(\begin{array}{ccc}
\lambda & 0 & 1 \\
-\lambda \hat{a} & 0 & 0 \\
0 & 0 & 0 
\end{array}
\right)~~~, 
\nonumber \\
\hat{q}(\lambda) \hat{H}_{\dot{1}} 
&=& 
\left( 
\begin{array}{cccc}
0 & 
\frac{\hat{h}_{11}\!-\! \hat{h}_{31}}{\lambda} & 
\cdots & 
\frac{\hat{h}_{1 n\!-\!1}
\!-\! \hat{h}_{3 n\!-\!1}}{\lambda} 
\\ 
0 & 
\hat{h}_{21}\!-\! 
\frac{\hat{a}(\hat{h}_{11}
\!-\! \hat{h}_{31})}{\lambda} & 
\cdots & 
\hat{h}_{2 n\!-\!1}\!-\! 
\frac{\hat{a}
(\hat{h}_{1n\!-\!1}\!-\! 
\hat{h}_{3n\!-\!1})}{\lambda} 
\\
1 & \hat{h}_{31} & \cdots & \hat{h}_{3 n\!-\!1} 
\end{array}
\right)~~. 
\label{example2-3}
\end{eqnarray}
Rewritting (\ref{example2-3}) in terms of rescaled 
parameters (\ref{rescale1 example2}) and 
(\ref{rescale2 example2}) it exactly 
coincides with (\ref{example2-2}) at 
$P=Q$.

\subsubsection*{Example 3 : $(m,m')=(2,0)$}

            As the last example we shall 
investigate the case 
of four D5-branes two of which overlap 
at $P=(0,\lambda)$ 
and the other two are degenerate at $Q=(0,0)$. 
$\lambda \neq 0$. 
The corresponding vacuum 
$(\hat{B}_a,\hat{H}_{\dot{A}})$ 
has the form 
\begin{eqnarray}
&& 
\hat{B}_1= 
\left( \begin{array}{cccc}
0 & 1 & 0 & 0 \\
0 & 0 & 0 & 0 \\ 
0 & 0 & 0 & 1 \\ 
0 & 0 & 0 & 0 
\end{array}\right)~~,~~ 
\hat{B}_2= 
\left( \begin{array}{cccc}
\lambda & \hat{a} & 0 & 0 \\ 
0     & \lambda & 0 & 0 \\ 
0 & 0 & 0 & \hat{b} \\ 
0 & 0 & 0 & 0 
\end{array}\right)~~, 
\nonumber \\ 
&& 
\hat{H}_{\dot{1}}= 
\left( \begin{array}{cccc}
0 & \hat{h}_{11} & \cdots & \hat{h}_{1n\!-\!1} \\ 
1 & \hat{h}_{21} & \cdots & \hat{h}_{2n\!-\!1} \\ 
0 & \hat{h}_{31} & \cdots & \hat{h}_{3n\!-\!1} \\ 
1 & \hat{h}_{41} & \cdots & \hat{h}_{4n\!-\!1} 
\end{array}\right)~~,~~ 
\hat{H}_{\dot{2}}=0~~, 
\label{example3-1} 
\end{eqnarray}
where $\hat{a},\hat{b}$ and $\hat{h}_{ij}$ are  
$4n-2$ complex parameters in which, 
taking the diagonal parts of $\hat{B}_a$ into account, 
$\hat{a}$,$\hat{h}_{1j}$ and $\hat{h}_{2j}$ 
are the additional parameters associated with 
the two D5-branes degenerate at $P$ 
while $\hat{b}$, $\hat{h}_{3j}$ and $\hat{h}_{4j}$ are 
those for the two D5-branes at $Q$. 
It is easy to see that 
$\hat{B}_{a}$ in (\ref{example3-1}) satisfy the relations, 
\begin{eqnarray}
&& 
\hat{B}_1^2=0~~, 
\nonumber \\ 
&& 
\left(\hat{B}_2-\hat{a}\hat{B}_1-\lambda \right) 
\left(\hat{B}_2-\hat{b}\hat{B}_1\right)=0~~, 
\nonumber \\  
&& 
\hat{B}_1\hat{B}_2=\hat{B}_2\hat{B}_1~~. 
\label{eq of B example3-1} 
\end{eqnarray}
Since $\hat{H}_{\dot{1}}(e_1)$ is cyclic with respect to 
$\hat{B}_a$ one can write down $\hat{H}_{\dot{1}}(e_j)$ 
$(j \geq 2)$ in the form, 
\begin{eqnarray} 
\hat{H}_{\dot{1}}(e_j) &=&  
\left\{ 
\left( 
\frac{\hat{h}_{1j}-\hat{h}_{3j}}{\lambda}
       -\frac{(\hat{a}-\hat{b})
           (\hat{h}_{2j}-\hat{h}_{4j})}{\lambda^2}
\right) 
\hat{B}_1\hat{B}_2+ 
\left( 
\hat{h}_{3j}-
\frac{\hat{b}(\hat{h}_{2j}-\hat{h}_{4j})}{\lambda} 
\right) 
\hat{B}_1 
\right. 
\nonumber \\ 
&& ~~~~~~~~~~~~~~~~~~~~~
\left.+ 
\frac{\hat{h}_{2j}-\hat{h}_{4j}}{\lambda}\hat{B}_2+
\hat{h}_{4j} 
\right\} 
\hat{H}_{\dot{1}}(e_1)~~.
\end{eqnarray}
The corresponding ``gauge transform" $\hat{G}$, 
which has a similar form as 
(\ref{gauge example1-1}), 
can be described by the following polynomials 
$\hat{g}_j(z)$ $(j \geq 2)$: 
\begin{eqnarray} 
\hat{g}_{j}(z) &=&  
\left( 
\frac{\hat{h}_{1j}-\hat{h}_{3j}}{\lambda}
       -\frac{(\hat{a}-\hat{b})
                (\hat{h}_{2j}-\hat{h}_{4j})}
                    {\lambda^2}
\right) 
z_1z_2+ 
\left( 
\hat{h}_{3j}-
\frac{\hat{b}(\hat{h}_{2j}-\hat{h}_{4j})}{\lambda} 
\right) 
z_1  
\nonumber \\ 
&& ~~~~~~~~~~~~~~~~~~~~~~~
+ 
\frac{\hat{h}_{2j}-\hat{h}_{4j}}{\lambda}z_2+
\hat{h}_{4j}~~. 
\label{hatg example3}
\end{eqnarray}

           As regards the configuration obtained from 
(\ref{example3-1}) by setting $\lambda =0$, 
the first column 
vector of $\hat{H}_{\dot{1}}$ is not cyclic. 
For this configuration to be a vacuum of D5-branes 
it may be enough, say, if, 
in addition to the first column 
vector $\hat{H}_{\dot{1}}(e_1)$ and its descendant 
$\hat{B}_1\hat{H}_{\dot{1}}(e_1)$, 
the second column vector $\hat{H}_{\dot{1}}(e_2)$ and 
its descendant $\hat{B}_1\hat{H}_{\dot{1}}(e_2)$ span 
${\bf C}^4$ ($\equiv V$). This condition leads the 
constraint, $\hat{h}_{21} \neq \hat{h}_{41}$. 
If it is satisfied, 
one can introduce the new bases of $V$ with which 
vacuum (\ref{example3-1}) (with setting $\lambda =0$) 
acquires the form 
\begin{eqnarray} 
&& 
\hat{B}_1= 
\left( \begin{array}{cccc}
0 & 0 & 1 & \hat{h}_{21} \\
0 & 0 & 0 & 1 \\ 
0 & 0 & 0 & 0 \\ 
0 & 0 & 0 & 0 
\end{array}\right)~~,~~ 
\hat{B}_2= 
\left( \begin{array}{cccc}
0 & 0 & \hat{b} & \hat{b}\hat{h}_{21}\!-\! 
      \frac{\hat{a}\!-\! \hat{b}}
             {\hat{h}_{21}\!-\! \hat{h}_{41}} \\ 
0     & 0 & 0 & \hat{a} \\ 
0 & 0 & 0 & \hat{b} \\ 
0 & 0 & 0 & 0 
\end{array}\right)~~, 
\nonumber \\ 
&& 
\hat{H}_{\dot{1}}= 
\left( \begin{array}{cccc}
0 & 0 & * & * \\ 
0 & 0 & * & * \\ 
0 & 1 & * & * \\ 
1 & 0 & * & * 
\end{array}\right)~~,~~ 
\hat{H}_{\dot{2}}=0~~. 
\end{eqnarray} 
This means that parameters $\hat{h}_{11,31}$ 
are absorbed into the $GL(4:{\bf C})$-symmetry.

          Nextly let us describe the vacuum 
$(B_a,H_{\dot{A}})$ which belongs to the maximal cycle 
$C_{({\scriptsize \mbox{[2,2]}},
\emptyset,\cdots,\emptyset)}$. 
It has the realization : 
\begin{eqnarray} 
&& 
B_1= 
\left( \begin{array}{cccc}
0 & 1 & 0 & 0 \\
0 & 0 & 0 & 0 \\ 
0 & 0 & 0 & 1 \\ 
0 & 0 & 0 & 0 
\end{array}\right)~~,~~ 
B_2= 
\left( \begin{array}{cccc}
0 & a & 1 & 0 \\ 
0 & 0 & 0 & 1 \\ 
0 & b & 0 & a \\ 
0 & 0 & 0 & 0 
\end{array}\right)~~, 
\nonumber \\ 
&& 
H_{\dot{1}}= 
\left( \begin{array}{cccc}
0 & h_{11} & \cdots & h_{1n\!-\!1} \\ 
0 & h_{21} & \cdots & h_{2n\!-\!1} \\ 
0 & h_{31} & \cdots & h_{3n\!-\!1} \\ 
1 & h_{41} & \cdots & h_{4n\!-\!1} 
\end{array}\right)~~,~~ 
H_{\dot{2}}=0~~, 
\label{example3-2}
\end{eqnarray}
where complex parameters $a,b$ and $h_{ij}$ 
constitute the maximal cycle. 
$B_a$ in (\ref{example3-2}) satisfy the relations, 
\begin{eqnarray} 
&&
B_1^2=0~~, 
\nonumber \\ 
&& 
B_2^2-2aB_1B_2-bB_1=0~~, 
\nonumber \\ 
&& 
B_1B_2=B_2B_1~~. 
\label{eq of B example3-2} 
\end{eqnarray} 
The elements of the ``gauge transform" $G$ 
in (\ref{gauge}) is now given by 
\begin{eqnarray}
g_j(z)=
h_{1j}z_1z_2+(h_{3j}-ah_{2j})z_1+
h_{2j}z_2+h_{4j}~~. 
\label{g example3-2}
\end{eqnarray}

           The behavior of 
configuration (\ref{example3-1}) 
when $P$ goes to $Q$ can be studied 
by appropriate rescalings of the additional parameters. 
Let us first rescale $\hat{a}$ and $\hat{b}$ in 
(\ref{example3-1}) such that 
$\lambda(\hat{a}-\hat{b})$ and $\hat{a}+\hat{b}$ do not 
depend on $\lambda$. Introduce the new parameters by 
\begin{eqnarray} 
a&=& 
\frac{\hat{a}+\hat{b}}{2}~~, 
\nonumber \\ 
b&=& 
\frac{\lambda(\hat{a}-\hat{b})}{2}~~. 
\label{rescale1 example3} 
\end{eqnarray} 
With fixing them one can see  
relations (\ref{eq of B example3-1}) become 
(\ref{eq of B example3-2}) as $\lambda$ goes to zero. 
As regards the other additional parameters 
in (\ref{example3-1}) 
we shall rescale them such that 
\begin{eqnarray} 
&& 
h_{1j}=
\frac{\hat{h}_{1j}-\hat{h}_{3j}}{\lambda}- 
\frac{(\hat{a}-\hat{b})(\hat{h}_{2j}-\hat{h}_{4j})}
          {\lambda^2}~~, 
\nonumber \\ 
&& 
h_{2j}= 
\frac{\hat{h}_{2j}-\hat{h}_{4j}}{\lambda}~~, 
\nonumber \\ 
&& 
h_{3j}= 
\hat{h}_{3j}+ 
\frac{(\hat{a}-\hat{b})(\hat{h}_{2j}-\hat{h}_{4j})}
            {2\lambda}~~, 
\nonumber \\ 
&& 
h_{4j}= \hat{h}_{4j} ~~
\label{rescale2 example3}
\end{eqnarray} 
become independent of $\hat{a},\hat{b}$ and $\lambda$. 
The elements $\hat{g}_j(z)$ of the gauge transform 
$\hat{G}$, 
expressing them in terms of these rescaled parameters, 
coincide with (\ref{g example3-2}). 
Notice that equations (\ref{rescale1 example3}) and 
(\ref{rescale2 example3}) can be also regarded as 
the change of coordinates as far as $\lambda \neq 0$. 
Introduce the element 
$\hat{q}(\lambda)$ of $GL(4:{\bf C})$ 
\begin{eqnarray}
\hat{q}(\lambda)= 
\left( \begin{array}{cccc} 
1/\lambda & -(\hat{a}-\hat{b})/\lambda^2 & 
-1/\lambda & (\hat{a}-\hat{b})\lambda^2 \\ 
0 & 1/\lambda & 0 & -1/\lambda \\ 
0 & (\hat{a}-\hat{b})/2\lambda & 1 & 
-(\hat{a}-\hat{b})/2\lambda \\ 
0 & 0 & 0 & 1 
\end{array}\right)~~~.
\end{eqnarray} 
The transform of (\ref{example3-1}) 
by $\hat{q}(\lambda)$ 
becomes 
\begin{eqnarray} 
\hat{q}(\lambda)\hat{B}_1\hat{q}(\lambda)^{-1}
&=& 
\left( \begin{array}{cccc} 
0 & 1 & 0 & 0 \\ 
0 & 0 & 0 & 0 \\ 
0 & 0 & 0 & 1 \\ 
0 & 0 & 0 & 0 
\end{array} \right) ~~,~~ 
\nonumber \\ 
\hat{q}(\lambda)\hat{B}_2\hat{q}(\lambda)^{-1} 
&=& 
\left( \begin{array}{cccc} 
\lambda & (\hat{a}+\hat{b})/2 & 1 & 0 \\ 
0 & \lambda & 0 & 1 \\ 
0 & \lambda(\hat{a}-\hat{b})/2 & 0 & 
(\hat{a}+\hat{b})/2 \\ 
0 & 0 & 0 & 0 
\end{array} \right) ~~,~~ 
\nonumber \\ 
\hat{q}(\lambda)\hat{H}_{\dot{1}} 
&=&
\left(\begin{array}{cccc} 
0 & 
\frac{\hat{h}_{11}-\hat{h}_{31}}{\lambda}- 
\frac{(\hat{a}-\hat{b})(\hat{h}_{21}-\hat{h}_{41})}
{\lambda^2} & 
\cdots & 
\frac{\hat{h}_{1n\!-\!1}-\hat{h}_{3n\!-\!1}}
        {\lambda}- 
\frac{(\hat{a}-\hat{b})
       (\hat{h}_{2n\!-\!1}-\hat{h}_{4n\!-\!1})}
        {\lambda^2} 
\\ 
0 & 
\frac{\hat{h}_{21}-\hat{h}_{41}}{\lambda} & 
\cdots & 
\frac{\hat{h}_{2n\!-\!1}-\hat{h}_{4n\!-\!1}}
        {\lambda} 
\\  
0 & 
\hat{h}_{31}+ 
\frac{(\hat{a}-\hat{b})(\hat{h}_{21}-\hat{h}_{41})}
        {2\lambda} & 
\cdots & 
\hat{h}_{3n\!-\!1}+ 
\frac{(\hat{a}-\hat{b})
  (\hat{h}_{2n\!-\!1}-\hat{h}_{4n\!-\!1})}{2\lambda}    
\\ 
1 &  
\hat{h}_{41} & 
\cdots & 
\hat{h}_{4n\!-\!1} 
\end{array}\right)~~.
\nonumber \\
&&~~~~~
\label{example3-3} 
\end{eqnarray} 
Rewritting (\ref{example3-3}) 
in terms of the rescaled parameters 
we can see that it becomes (\ref{example3-2}) 
as $P$ goes to $Q$.

\sect{Toward Second-Quantization of D5-Brane}
\subsection*{Superposition of D5-Brane Vacua}

               In the previous section 
it is shown that the additional complex parameters 
of vacuum (\ref{cycle2}) of $k=k_1+k_2+\cdots+k_l$ 
D5-branes form a cycle in the moduli space 
${\cal M}(k)$ which can be 
identified with the maxiaml cycle 
$C_{(\Gamma,\emptyset,\cdots,\emptyset)}$ 
with $\Gamma=\mbox{$[k_1,k_2,\cdots,k_l]$}$. 
Vacuum (\ref{cycle2}) is described 
by the superposition of vacua of 
$k_i$ pieces ($1 \leq i \leq l$). 
Each constituent, that is, 
vacuum (\ref{element of cycle2}) 
of $k_i$ D5-branes degenerate at $P_i$, 
admits the additional degrees of freedom, 
which contributes to the parameters 
of (\ref{cycle2}). 
Even when some of the positions of these 
overlapping D5-branes coincide, say, 
$P_i=P_j$, 
this superposed vacuum of $k$ D5-branes 
is shown to be still well-defined 
with an approriate change of realization. 
In order to make the positions of 
these overlapping D5-branes explicit 
let us write the above topological cycle by 
$C_{{\scriptsize \mbox{$[k_1,k_2,\cdots,k_l]$}}}
(P_1,P_2,\cdots,P_l)$ 
where 
$P_i=(z_1^{(i)},z_2^{(i)})$ is the position of  
$k_i$ coincident D5-branes. 
$C_{{\scriptsize \mbox{$[k_1,\cdots,k_l]$}}}
(P_1,\cdots,P_l) 
\simeq C_{({\scriptsize \mbox{$[k_1,\cdots,k_l]$}},
\emptyset,\cdots,\emptyset)}$.

         It is physically reasonable to consider  
such a vacuum of $k+\tilde{k}$ D5-branes 
in which the configuration of $k$ pieces  
can be regarded as the vacuum given by 
(\ref{cycle2}) 
while the configuration of the other 
$\tilde{k}$ pieces is an arbitrary vacuum 
of $\tilde{k}$ D5-branes. 
Taking it in the reverse order one might say  
it should be possible, 
at least from the viewpoint of 
five-branes, to construct a new vacuum 
of $k+\tilde{k}$ D5-branes 
from two vacua of $k$ and $\tilde{k}$ pieces. 
In spite of this physical speculation 
we should note that the superposition 
of two vacua, 
though it works directly when both vacua 
of $k$ and $\tilde{k}$ D5-branes 
admit to have forms similar to (\ref{cycle2}), 
does not work directly 
if either vacuum of $k$ or $\tilde{k}$ 
pieces is generic  
\footnote{We can easily superpose two vacua 
in the case of $n=1$. This is because, when $n=1$, 
one can set \cite{Nakajima} $H_{\dot{2}}=0$ 
for all the vacua $(B_a,H_{\dot{A}})$.}. 
So, it is necessary to explain 
how one can ``superpose" the vacuum of $k$ D5-branes  
which belongs to the cycle 
$C_{{\scriptsize \mbox{$[k_1,\cdots,k_l]$}}}
(P_1,\cdots,P_l)$ 
with a vacuum of $\tilde{k}$ D5-branes 
without restricting the latter.

               Let $(B_a^{(1)},H_{\dot{A}}^{(1)})$  
be the vacuum of $k$ D5-branes which has 
form (\ref{cycle2}). It belongs to the cycle 
$C_{{\scriptsize \mbox{$[k_1,\cdots,k_l]$}}}
(P_1,\cdots,P_l)$. 
Take a generic vacuum of $\tilde{k}$ D5-branes, 
which is denoted by $(B_a^{(2)},H_{\dot{A}}^{(2)})$. 
It may be convenient to handle these vacua by using 
hyperk\"ahler description (\ref{moduli}) 
($or$ (\ref{ADHM1}) and (\ref{ADHM2})) of the moduli 
spaces. Notice that there exist 
$g(\eta)^{(1)} \in GL(k:{\bf C})$ 
and    
$g(\eta)^{(2)} \in GL(\tilde{k}:{\bf C})$ 
such that the transforms of 
$(B_a^{(1)},H_{\dot{A}}^{(1)})$ 
by $g^{(1)}(\eta)$ 
and 
$(B_a^{(2)},H_{\dot{A}}^{(2)})$ 
by $g^{(2)}(\eta)$ 
both satisfy $D$-flat conditions 
(\ref{ADHM1}) and (\ref{ADHM2}). 
These transforms will be denoted by 
$(B_a^{(1,2)}(\eta),H_{\dot{A}}^{(1,2)}(\eta))$  
in order to make their dependence on 
$\eta$ explicit.
For the same reason 
the corresponding moduli spaces, 
regarded as the hyperk\"ahler quotients, 
will be denoted by   
${\cal M}(k)_{\eta}$ and 
${\cal M}(\tilde{k})_{\eta}$. 
Notice that  
$(B_a^{(1)}(\eta),H_{\dot{A}}^{(1)}(\eta)) 
\in {\cal M}(k)_{\eta}$ 
and 
$(B_a^{(2)}(\eta),H_{\dot{A}}^{(2)}(\eta))  
\in {\cal M}(\tilde{k})_{\eta} $. 
We first consider their behaviors 
as $\eta$ goes to zero. 
For the case of $\eta$ being positive  
the vacuum 
$(B_a^{(1)}(\eta),H_{\dot{A}}^{(1)}(\eta))$ 
admits additional parameters 
as it has in the complex symplectic description. 
But, 
as $\eta$ goes to zero, 
these additional degrees of freedom vanish.  
It will become 
\begin{eqnarray}
B_a^{(1)}(0) 
&\equiv& 
\lim_{\eta \rightarrow +0}B_a^{(1)}(\eta) 
\nonumber \\ 
&=& 
\left( \begin{array}{cccc} 
z_a^{(1)}{\bf 1}_{k_1} & 0 & \cdots & 0 \\ 
0 & \ddots & \ddots & \vdots \\ 
\vdots & \ddots & \ddots & 0 \\ 
0 & \cdots & 0 & z_a^{(l)}{\bf 1}_{k_l} 
\end{array} \right)~~, 
\nonumber \\ 
H_{\dot{A}}^{(1)}(0) 
&\equiv& 
\lim_{\eta \rightarrow +0}H_{\dot{A}}^{(1)}(\eta) 
\nonumber \\ 
&=& 
0~~.
\label{small instanton} 
\end{eqnarray} 
Notice that 
$P_i=(z_1^{(i)},z_2^{(i)})$ $(1 \leq i \leq l)$ 
are the positions of overlapping $k_i$ D5-branes.  
Configuration (\ref{small instanton}) 
can be regarded as a vacuum 
at $\eta =0$.  
In the terminology of 
four-dimensional gauge theory, 
it describes the small size limit of 
$k$ $SU(n)$-instantons.  
$k_i$ of these ideal instantons 
are located at $P_i$.  
As regards the vacuum 
$(B_a^{(2)}(0),H_{\dot{A}}^{(2)}(0))$ 
$\equiv \lim_{\eta \rightarrow +0}
(B_a^{(2)}(\eta),H_{\dot{A}}^{(2)}(\eta))$, 
since it is a generic point of the moduli space 
of $\tilde{k}$ D5-branes, it describes  
non-singular $\tilde{k}$ $SU(n)$-instantons. 
At $\eta =0$ one can easily 
superpose these two vacua. 
Namely let us introduce 
\begin{eqnarray} 
B_a(0) \equiv 
\left( \begin{array}{cc} 
B_a^{(1)}(0) & 0 \\ 
0 & B_a^{(2)}(0) 
\end{array} \right)~~,~~ 
H_{\dot{A}}(0) \equiv  
\left( \begin{array}{c} 
H_{\dot{A}}^{(1)}(0) \\ 
H_{\dot{A}}^{(2)}(0) 
\end{array}\right)~~. 
\label{superpose at 0}
\end{eqnarray} 
Then configuration (\ref{superpose at 0}) 
can be regarded as a point 
of the moduli space 
${\cal M}(k+\tilde{k})_{\eta =0}$. 
The contribution of the first $k$ D5-branes 
makes it a singular point.

           It is convenient to 
remark on the relation between the moduli spaces 
${\cal M}(k+\tilde{k})_{\eta >0}$ and 
${\cal M}(k+\tilde{k})_{\eta =0}$. 
As was explained in Section.2,  
the moduli space ${\cal M}(k+\tilde{k})_{\eta >0}$ 
is smooth. It has no singularity. 
Under the limit that $\eta$ goes to zero 
singularities do appear.  
These singularities correspond 
to the small size limits of $SU(n)$-insatntons. 
This phenomenon shows that 
{\it ${\cal M}(k+\tilde{k})_{\eta >0}$ 
can be regarded as a resolution of 
${\cal M}(k+\tilde{k})_{\eta =0}$.} 
The projection $\pi$ 
from ${\cal M}(k+\tilde{k})_{\eta >0}$ 
to ${\cal M}(k+\tilde{k})_{\eta =0}$ 
is defined by taking the limit of $\eta$ being zero 
in the former moduli space.  
As regards configuration (\ref{superpose at 0}), 
which is a singular point of 
${\cal M}(k+\tilde{k})_{\eta =0}$,  
the inverse image of (\ref{superpose at 0}), 
that is,  
$\pi^{-1}((B_a(0),H_{\dot{A}}(0)))$ 
gives us the vacua 
which admit additional complex parameters 
by which the singularity is resolved. 
Among these vacua there exists a configuration 
$(B_a(\eta),H_{\dot{A}}(\eta))$ 
which corresponds to a superposition 
of the two vacua 
$(B_a^{(1)}(\eta),H_{\dot{A}}^{(1)}(\eta))$ 
and 
$(B_a^{(2)}(\eta),H_{\dot{A}}^{(2)}(\eta))$. 
It realizes the ``superposition" of the vacuum of 
$k$ D5-branes which belongs to the cycle 
$C_{{\scriptsize \mbox{$[k_1,\cdots,k_l]$}}}
(P_1,\cdots,P_l)$ 
and a generic vacuum of $\tilde{k}$ D5-branes.

         Let us comment briefly on the case of 
$(B_a^{(2)},H_{\dot{A}}^{(2)})$ being a vacuum 
which causes a singularity at $\eta=0$. 
Suppose  
$(B_a^{(2)},H_{\dot{A}}^{(2)})$ 
$\in C_{\tilde{\Gamma}}
(\tilde{P}_1,\cdots,\tilde{P}_{\tilde{l}})$. 
($|\tilde{\Gamma}|=\tilde{k}$ and 
$l(\tilde{\Gamma})=\tilde{l}$.) 
When all the $\tilde{l}$-positions 
$\tilde{P}_1,\cdots,\tilde{P}_{\tilde{l}}$ 
are different from the $l$-positions 
$P_1,\cdots,P_l$ of the first $k$ D5-branes, 
their superposition will be given 
\footnote{
Here we use the complex symplectic description.} 
\begin{eqnarray}
B_a \equiv 
\left( \begin{array}{cc} 
B_a^{(1)} & 0 \\ 
0 & B_a^{(2)} 
\end{array} \right)~~,~~ 
H_{\dot{A}} \equiv 
\left( \begin{array}{c} 
H_{\dot{A}}^{(1)} \\ 
H_{\dot{A}}^{(2)} 
\end{array} \right)~~. 
\label{superpose} 
\end{eqnarray}
Clearly configuration (\ref{superpose}) 
belongs to the cycle 
$C_{\Gamma \cup \tilde{\Gamma}}
(Q_1,\cdots,Q_{l+\tilde{l}})$ 
of the moduli space ${\cal M}(k+\tilde{k})$ 
where 
$\Gamma \cup \tilde{\Gamma}$ is the Young tableau 
of $k+\tilde{k}$ boxes obtained from  
$\Gamma=\mbox{$[k_1,\cdots,k_l]$}$ 
and 
$\tilde{\Gamma}
=\mbox{$[\tilde{k}_1,\cdots,
\tilde{k}_{\tilde{l}}]$}$ 
by reordering $k_i$ and $\tilde{k}_j$. 
$Q_1,\cdots,Q_{l+\tilde{l}}$ 
are the corresponding rearrangement of 
$P_1,\cdots,P_l$ 
and 
$\tilde{P}_1,\cdots,\tilde{P}_{\tilde{l}}$.  
In the case that some of the positions of 
$\tilde{k}$ D5-branes 
coincide with those of the first 
$k$ D5-branes, 
owing to the discussion given in 
the previous section, 
it is still possible to give 
their superposition in the cycle 
$C_{\Gamma \cup \tilde{\Gamma}}
(Q_1,\cdots,Q_{l+\tilde{l}})$.

           The ``superpostion" defines 
an inclusion $\iota$ of 
$C_{{\scriptsize \mbox{$[k_1,\cdots,k_l]$}}}
(P_1,\cdots,P_l) 
\times {\cal M}(\tilde{k})$ 
to ${\cal M}(k+\tilde{k})$ : 
\begin{eqnarray}
\iota~~~~~
~~C_{{\scriptsize \mbox{$[k_1,\cdots,k_l]$}}}
         (P_1,\cdots,P_l) 
\times {\cal M}(\tilde{k}) 
~\hookrightarrow~
{\cal M}(k+\tilde{k})~~~.  
\label{inclusion map}
\end{eqnarray}
The image of $\iota$, that is, 
$\iota $
$\left(C_{{\scriptsize \mbox{$[k_1,\cdots,k_l]$}}}
(P_1,\cdots,P_l) 
\times {\cal M}(\tilde{k}) \right)$,  
is a noncompact submanifold of the moduli space 
${\cal M}(k+\tilde{k})$, which will be denoted by  
${\cal C}_{{\scriptsize \mbox{$[k_1,\cdots,k_l]$}}}
(P_1,$$\cdots,P_l)$. 
The dimensions of 
${\cal C}_{{\scriptsize \mbox{$[k_1,\cdots,k_l]$}}}
(P_1,$$\cdots,P_l)$ is 
equal to the sum of the dimensions of 
the moduli space ${\cal M}(\tilde{k})$ and 
the cycle 
$C_{{\scriptsize \mbox{$[k_1,\cdots,k_l]$}}}
(P_1,$$\cdots,P_l)$,  
\begin{eqnarray}
dim~ 
{\cal C}_{{\scriptsize \mbox{$[k_1,\cdots,k_l]$}}}
(P_1,\cdots,P_l)
=4n \tilde{k}+2(nk-l). 
\label{dim of submfd C(P)}
\end{eqnarray} 
Notice that $k_1+\cdots+k_l=k$. 
Physically speaking,  
any point of the submanifold 
${\cal C}_{{\scriptsize \mbox{$[k_1,\cdots,k_l]$}}}
(P_1,$$\cdots,P_l)$ 
describes the vacuum of $k+\tilde{k}$ D5-branes 
in which the configuration of $k$ pieces, 
considering it as a vacuum of $k$ D5-branes, 
belongs to the cycle 
$C_{{\scriptsize \mbox{$[k_1,\cdots,k_l]$}}}
(P_1,$$\cdots,P_l)$ 
of the moduli space ${\cal M}(k)$. 
Each point $P_i$ denotes the position 
of overlapping $k_i$ D5-branes.   
By letting these $l$-positions 
of $k=k_1+k_2+\cdots+k_l$ 
D5-branes free in the four-dimensions, 
we may introduce the noncompact 
submanifold 
${\cal C}_{{\scriptsize \mbox{$[k_1,\cdots,k_l]$}}}$ 
of 
${\cal M}(k+\tilde{k})$ 
by 
\begin{eqnarray}
{\cal C}_{{\scriptsize \mbox{$[k_1,\cdots,k_l]$}}}
\equiv \left. \left\{
~{\cal C}_{{\scriptsize \mbox{$[k_1,\cdots,k_l]$}}}
(P_1,\cdots,P_l)
~\right|
~P_1,\cdots,P_l \in {\bf C}^2
~\right\}~~. 
\label{submfd C}
\end{eqnarray}
The dimensions of this submanifold is 
\begin{eqnarray}
dim~ 
{\cal C}_{{\scriptsize \mbox{$[k_1,\cdots,k_l]$}}}
=4n\tilde{k}+2(nk+l). 
\label{dim of submfd C}
\end{eqnarray}

            Now let us exchange the roles of $k$ and 
$\tilde{k}$ D5-branes in the construction of 
inclusion map (\ref{inclusion map}). 
Consider the topological cycle 
$C_{\tilde{\Gamma}}
(\tilde{P}_1,$$\cdots,\tilde{P}_{\tilde{l}})$ 
of the moduli space ${\cal M}(\tilde{k})$. 
$\tilde{\Gamma}
=\mbox{$[\tilde{k}_1,\cdots,\tilde{k}_{\tilde{l}}]$}$
with $|\tilde{\Gamma}|=\tilde{k}$. 
Each point $\tilde{P}_i$ denotes the position of  
overlapping $\tilde{k}_i$ D5-branes. 
By the superposition of vacua we will obtain 
the inclusion map,   
\begin{eqnarray}
\tilde{\iota}~~~~~ 
~~{\cal M}(k) \times 
C_{{\scriptsize 
   \mbox{$[\tilde{k}_1,\cdots,\tilde{k}_{\tilde{l}}]$}}}
(\tilde{P}_1,\cdots,\tilde{P}_{\tilde{l}})  
~\hookrightarrow~
{\cal M}(k+\tilde{k})~~~,  
\label{inclusion map2}
\end{eqnarray}
which provides, as its image, 
the noncompact submanifold  
${\cal C}_{{\scriptsize 
\mbox{$[\tilde{k}_1,\cdots,\tilde{k}_{\tilde{l}}]$}}}
(\tilde{P}_1,\cdots,\tilde{P}_{\tilde{l}})$
of the moduli space ${\cal M}(k+\tilde{k})$. 
Letting the $\tilde{l}$-points 
$\tilde{P}_1,\cdots,\tilde{P}_{\tilde{l}}$ 
free we also obtain the noncompact submanifold 
${\cal C}_{{\scriptsize 
  \mbox{$[\tilde{k}_1,\cdots,\tilde{k}_{\tilde{l}}]$}}}$. 
The dimensions of this submanifold is 
equal to $4nk+2(n\tilde{k}+\tilde{l})$. 
Notice that any point of 
the submanifold ${\cal C}_{{\scriptsize 
  \mbox{$[\tilde{k}_1,\cdots,\tilde{k}_{\tilde{l}}]$}}}$ 
describes the vacuum of $k+\tilde{k}$ D5-branes 
in which the configuration of $\tilde{k}$ pieces, 
considering it as a vacuum of $\tilde{k}$ D5-branes, 
belongs to the cycle 
$C_{{\scriptsize
  \mbox{$[\tilde{k}_1,\cdots,\tilde{k}_{\tilde{l}}]$}}}
(\tilde{P}_1,$$\cdots,\tilde{P}_{\tilde{l}})$ 
of the moduli space ${\cal M}(\tilde{k})$. 
By examining two vacua of $k+\tilde{k}$ 
D5-branes which respectively belong to  
${\cal C}_{{\scriptsize \mbox{$[k_1,\cdots,k_l]$}}}$ 
and 
${\cal C}_{{\scriptsize 
 \mbox{$[\tilde{k}_1,\cdots,\tilde{k}_{\tilde{l}}]$}}}$ 
one can find that 
their intersection  
in the moduli space ${\cal M}(k+\tilde{k})$ 
has the form, 
\begin{eqnarray}
{\cal C}_{{\scriptsize \mbox{$[k_1,\cdots,k_l]$}}} 
~\cap~ 
{\cal C}_{{\scriptsize 
   \mbox{$[\tilde{k}_1,\cdots,\tilde{k}_{\tilde{l}}]$}}}~=~ 
{\cal C}_{{\scriptsize \mbox{$[k_1,\cdots,k_l]$}} \cup 
{\scriptsize \mbox{$[\tilde{k}_1,
           \cdots,\tilde{k}_{\tilde{l}}]$}}}~~. 
\label{intersection 1}
\end{eqnarray}
The noncompact submanifold 
${\cal C}_{{\scriptsize \mbox{$[k_1,\cdots,k_l]$}} \cup 
{\scriptsize 
\mbox{$[\tilde{k}_1,\cdots,\tilde{k}_{\tilde{l}}]$}}}$ 
is simply realized as the set of the cycles 
$C_{{\scriptsize \mbox{$[k_1,\cdots,k_l]$}} \cup 
{\scriptsize \mbox{$[\tilde{k}_1,
\cdots,\tilde{k}_{\tilde{l}}]$}}}
(Q_1,\cdots,Q_{l+\tilde{l}})$ 
of the moduli space ${\cal M}(k+\tilde{k})$ : 
\begin{eqnarray}
{\cal C}_{{\scriptsize 
\mbox{$[k_1,\cdots,k_l]$}} \cup 
{\scriptsize \mbox{$[\tilde{k}_1,
\cdots,\tilde{k}_{\tilde{l}}]$}}}
&=& 
\left.\left\{ 
C_{{\scriptsize \mbox{$[k_1,\cdots,k_l]$}} \cup 
{\scriptsize \mbox{$[\tilde{k}_1,
\cdots,\tilde{k}_{\tilde{l}}]$}}}
(Q_1,\cdots,Q_{l+\tilde{l}})~\right|~ 
Q_1,\cdots,Q_{l+\tilde{l}} 
\in {\bf C}^2 \right\}~~~.
\nonumber \\ 
&&~~~~~~~
\end{eqnarray}  
Since 
$C_{{\scriptsize \mbox{$[k_1,\cdots,k_l]$}} \cup 
{\scriptsize \mbox{$[\tilde{k}_1,
\cdots,\tilde{k}_{\tilde{l}}]$}}}
(Q_1,\cdots,Q_{l+\tilde{l}})$ 
is isomorphic to  
$C_{({\scriptsize \mbox{$[k_1,\cdots,k_l]$}} \cup 
{\scriptsize \mbox{$[\tilde{k}_1,
\cdots,\tilde{k}_{\tilde{l}}]$}},
\emptyset,\cdots,\emptyset)}$, 
we can identify the submanifold 
${\cal C}_{{\scriptsize 
\mbox{$[k_1,\cdots,k_l]$}} \cup 
{\scriptsize \mbox{$[\tilde{k}_1,
\cdots,\tilde{k}_{\tilde{l}}]$}}}$ 
with 
$\left( {\bf C}^2 \right)^{\oplus (l+\tilde{l})} 
\times 
C_{({\scriptsize \mbox{$[k_1,\cdots,k_l]$}} \cup 
{\scriptsize 
\mbox{$[\tilde{k}_1,\cdots,\tilde{k}_{\tilde{l}}]$}},
\emptyset,\cdots,\emptyset)}$. 
Here 
``$\left( {\bf C}^2 \right)^{\oplus (l+\tilde{l})}$" 
parametrize the $(l+\tilde{l})$-positions 
where $k+\tilde{k}$ D5-branes are overlapping.

            One can also generalize inclusion map 
(\ref{inclusion map}) to the following direction. 
For each $i$, consider the inclusion map 
\begin{eqnarray}
C_{{\scriptsize \mbox{$[k_i]$}}}(P_i)
\times 
{\cal M}(\tilde{k}+\sum_{j \neq i}^lk_j)~~ 
\hookrightarrow~~
{\cal M}(k+\tilde{k})~~,
\label{i-th inclusion}
\end{eqnarray}
which now gives the submanifold 
${\cal C}_{{\scriptsize \mbox{$[k_i]$}}}$ 
of the moduli space ${\cal M}(k+\tilde{k})$. 
The dimensions of this submanifold is 
\begin{eqnarray}
dim~{\cal C}_{{\scriptsize \mbox{$[k_i]$}}}=
4n(\tilde{k}+\sum_{j \neq i}^lk_i)
+2(nk_i+1). 
\label{dim of Ci} 
\end{eqnarray}
The intersection of these submanifolds clearly turns out 
to be (\ref{submfd C}) :  
\begin{eqnarray}
{\cal C}_{{\scriptsize \mbox{$[k_1]$}}} \cap 
{\cal C}_{{\scriptsize \mbox{$[k_2]$}}} \cap \cdots \cap 
{\cal C}_{{\scriptsize \mbox{$[k_l]$}}}
={\cal C}_{{\scriptsize \mbox{$[k_1,\cdots,k_l]$}}=
         \cup_{i=1}^{l}{\scriptsize \mbox{$[k_i]$}}}~~.  
\label{intersection 2}
\end{eqnarray}

\subsection*{Physical Observables of Worldvolume Topological 
Field Theory} 

         Let us concentrate on the moduli space 
${\cal M}(k)$, that is, the classical vacua of 
$k$ D5-branes having open strings with the 
$U(n)$ Chan-Paton factors.  
In \cite{FKN} 
worldvolume topological $U(k)$ gauge theory has been 
constructed such that its physical content can be 
identified with cohomology theory of the moduli 
space ${\cal M}(k)$. 
In particular, 
the physical Hilbert space is realized by the 
cohomology group $H^*({\cal M}(k))$. 
In this subsection, taking this topological field 
theoretical viewpoint, 
we introduce a subclass of $H^*({\cal M}(k))$, which 
admits, 
as a class of the physical observables 
of topological field theory, 
to have a structure analogous to the Fock space.

                     We begin by considering 
the noncompact submanifold ${\cal C}_{\Gamma}$ 
with $|\Gamma|=k$ and $l(\Gamma)=l$. 
It is the set of the cycles 
$C_{\Gamma}(P_1,\cdots,P_l)$ with 
$P_i$ being arbitrary in the four-dimensions.  
Since each cycle $C_{\Gamma}(P_1,\cdots,P_l)$ 
can be topologically identified with 
the maximal cycle 
$C_{(\Gamma,\emptyset,\cdots,\emptyset)}$, 
it holds that 
${\cal C}_{\Gamma} \simeq 
\left( {\bf C}^2 \right)^{\oplus l} 
\times 
C_{(\Gamma, \emptyset,\cdots,\emptyset)}$. 
This isomorphism implies that 
the noncompact directions of the submanifold 
${\cal C}_{\Gamma}$ can be identified with  
``$\left( {\bf C}^2 \right)^{\oplus l}$",  
which simply parametrize the $l$-positions 
$P_1,\cdots,P_l$ where $k$ D5-branes overlap. 
Let us take the Poincar\'e dual of 
${\cal C}_{\Gamma}$ in 
the moduli space ${\cal M}(k)$. 
It will be denoted by 
${\cal O}_{\Gamma}$. 
The support of ${\cal O}_{\Gamma}$ is 
on the tubular neighborhood of 
${\cal C}_{\Gamma}$ and 
therefore it is noncompact. 
Taking account of the isomorphism 
${\cal C}_{\Gamma} \simeq$ 
$\left( {\bf C}^2 \right)^{\oplus l} 
\times C_{(\Gamma,\emptyset,\cdots,\emptyset)}$ 
one may modify ${\cal O}_{\Gamma}$ 
by adding or subtracting 
an appropriate exact form 
so that it does not depend 
on the noncompact directions. 
This means that 
one can make ${\cal O}_{\Gamma}$ 
independent of the coordinates of 
the $l$-positions 
$P_1,\cdots,P_l$ on $C_{\Gamma}(P_1,\cdots,P_l)$. 
(It will require 
a special care when some of these points 
coincide.) 
As regards the degrees of 
${\cal O}_{\Gamma}$ 
it is given by 
\begin{eqnarray} 
deg~ {\cal O}_{\Gamma} &=& 
dim~ {\cal M}(k)-~dim~{\cal C}_{\Gamma} 
\nonumber \\ 
&=& 
2(nk-l)~~, 
\label{dim of O} 
\end{eqnarray}
which is equal to 
the dimensions of the maximal cycle 
$C_{(\Gamma,\emptyset,\cdots,\emptyset)}$.

                       Suppose that 
$\Gamma=\mbox{$[k_1,\cdots,k_l]$}$. 
Consider the noncompact submanifold 
${\cal C}_{{\scriptsize \mbox{$[k_i]$}}}$ 
of the moduli space ${\cal M}(k)$. 
${\cal C}_{{\scriptsize \mbox{$[k_i]$}}}$ 
is the set of the vacua of $k$ D5-branes 
in which the configurations of $k_i$ pieces, 
considering it as vacua of 
$k_i$ five-branes, belong to the cycle 
$C_{{\scriptsize \mbox{$[k_i]$}}}(P)$ 
of the moduli space ${\cal M}(k_i)$. 
$P$ is arbitrary in the four-dimensions. 
Let us also take the Poincar\'e dual of 
this submanifold in the moduli space 
${\cal M}(k)$, 
which will be denoted by 
${\cal O}_{{\scriptsize \mbox{$[k_i]$}}}$. 
The support of 
${\cal O}_{{\scriptsize \mbox{$[k_i]$}}}$ 
is on the tubular neighbourhood of 
${\cal C}_{{\scriptsize \mbox{$[k_i]$}}}$. 
Through the inclusion map, 
$C_{{\scriptsize \mbox{$[k_i]$}}}(P)
\times {\cal M}(k-k_i)$
$\hookrightarrow {\cal M}(k)$, 
besides the identification of  
$C_{{\scriptsize \mbox{$[k_i]$}}}(P)$ 
with 
$C_{({\scriptsize \mbox{$[k_i]$}},
\emptyset,\cdots,\emptyset)}$ 
in the moduli space ${\cal M}(k_i)$, 
the submanifold 
${\cal C}_{{\scriptsize \mbox{$[k_i]$}}}$ 
can be regarded as 
$\left( {\bf C}^2 \right) \times {\cal M}(k-k_i) 
\times 
C_{({\scriptsize \mbox{$[k_i]$}},
\emptyset,\cdots,\emptyset)}$,  
where 
``${\bf C}^2$" 
parametrize the position of 
the overlapping $k_i$ D5-branes. 
Under this identification 
the noncompact directions of 
${\cal C}_{{\scriptsize \mbox{$[k_i]$}}}$ 
are 
$\left( {\bf C}^2 \right) \times {\cal M}(k-k_i)$. 
By adding or subtracting an appropriate exact form 
one may adjust 
${\cal O}_{{\scriptsize \mbox{$[k_i]$}}}$ 
such that 
it does not depend on these noncompact directions. 
Recalling 
$C_{({\scriptsize \mbox{$[k_i]$}},
\emptyset,\cdots,\emptyset)}$ 
is the cycle which is added in order to resolve 
the singularity of (overlapping) 
$k_i$ ideal $SU(n)$-instantons, 
this means that 
${\cal O}_{{\scriptsize \mbox{$[k_i]$}}}$ 
can be taken such that 
it only depends on the local 
data of the resolution. 
The degrees of 
${\cal O}_{{\scriptsize \mbox{$[k_i]$}}}$ 
turns out to be 
\begin{eqnarray} 
deg~{\cal O}_{{\scriptsize \mbox{$[k_i]$}}} &=& 
dim~{\cal M}(k)-
~dim~{\cal C}_{{\scriptsize \mbox{$[k_i]$}}} 
\nonumber \\ 
&=& 2(nk_i-1)~~,
\label{deg of Oi}
\end{eqnarray} 
which is independent of $k$ and equals to 
the dimensions of the cycle 
$C_{({\scriptsize \mbox{$[k_i]$}},
\emptyset,\cdots,\emptyset)}$ 
of the moduli space ${\cal M}(k_i)$.

             Now let us apply formula 
(\ref{intersection 2}). 
Taking the Poincar\'e dual of 
(\ref{intersection 2}) one can 
obtain the relation 
\begin{eqnarray} 
{\cal O}_{\Gamma=
      \cup_{i=1}^l{\scriptsize \mbox{$[k_i]$}}}
=
{\cal O}_{{\scriptsize \mbox{$[k_1]$}}} 
\wedge \cdots \wedge 
{\cal O}_{{\scriptsize \mbox{$[k_l]$}}}~~.
\label{co-intersection 2}
\end{eqnarray}
One of the implications of 
(\ref{co-intersection 2}) 
is as follows : 
As was explained in Section.2, 
several nontrivial cycles appear 
in the moduli space 
${\cal M}(k)$, 
which one may classify by using the set of 
$n$ Young tableaux 
$(\Gamma_1,\Gamma_2,\cdots,\Gamma_n)$ 
satisfying conditions (\ref{Young tableau}). 
Among them 
we have investigated the maximal cycles 
$C_{(\Gamma,\emptyset,\cdots,\emptyset)}$ 
with $|\Gamma|=k$. 
${\cal O}_{\Gamma}$ 
are introduced as the Poincar\'e duals 
of the submanifolds 
${\cal C}_{\Gamma}$ 
which can be identified with 
$\left( {\bf C}^2 \right)^{\otimes l(\Gamma)}
\times C_{(\Gamma,\emptyset,\cdots,\emptyset)}$. 
The noncompact degrees of freedom simply 
measure the positions 
where $k$ D5-branes are overlapping. 
One can regard the vector space spanned 
by these ${\cal O}_{\Gamma}$ 
as the cohomological version 
of the subspace of the homology group 
$H_* \left( {\cal M}(k) \right)$ 
spanned by the cycles 
$C_{(\Gamma,\emptyset,\cdots,\emptyset)}$. 
Formula (\ref{co-intersection 2}) 
implies that the subspace 
$\bigoplus_{|\Gamma|=k}{\bf C}{\cal O}_{\Gamma}$ 
of $H^*({\cal M}(k))$ 
admits a structure 
analogous to the Fock space  
\begin{eqnarray} 
\bigoplus_{|\Gamma|=k}
{\bf C}{\cal O}_{\Gamma} 
= 
\bigoplus_{m_1+\cdots+m_l=k}
{\bf C}{\cal O}_{m_1}\wedge \cdots \wedge 
{\cal O}_{m_l}~~, 
\label{restricted Fock space}
\end{eqnarray} 
where we simplify the notation as follows : 
\begin{eqnarray} 
{\cal O}_{m}\equiv 
{\cal O}_{{\scriptsize \mbox{$[m]$}}}~~. 
\end{eqnarray}

\subsection*{Second-Quantization of D5-Brane} 

            To recover the Fock space structure in 
(\ref{restricted Fock space}) it is enough to 
consider the direct sum  of these moduli spaces 
in stead of the specific moduli space.
This extension is quite reasonable 
from the viewpoint of five-branes 
because the number of five-branes does not suffer 
apriori any restriction. 
(Notice that we do not compactify the four-dimensions.) 
So, summing up $k$ in (\ref{restricted Fock space}) 
we obtain 
\begin{eqnarray}
{\cal H}_{total}\equiv 
\bigoplus_{l}
\bigoplus_{m_1,\cdots,m_l}
{\bf C}
{\cal O}_{m_1}\wedge \cdots \wedge 
{\cal O}_{m_l}~~,
\label{Fock space} 
\end{eqnarray} 
which is a subspace of 
$\bigoplus_k H^*({\cal M}(k))$. 
One might suspect that our definition of 
${\cal O}_m$ given in the previous subsection 
seems to depend on each moduli space 
${\cal M}(k)$ 
since it is introduced as the Poincar\'e dual 
of the submanifold 
${\cal C}_{{\scriptsize \mbox{$[m]$}}}$. 
But, by the closer look on  
the submanifold 
${\cal C}_{{\scriptsize \mbox{$[m]$}}}$, 
we can expect that ${\cal O}_m$ will be taken 
to depend only on the local data of the resolution 
of the singularity at $\eta=0$ 
caused by overlapping $m$ D5-branes 
and therefore it will acquire the form 
independent of $k$ \footnote{ 
This property is discussed 
in \cite{FKN} as the universality 
of the physical observables.}.

         We would like to propose that 
{\it ${\cal H}_{total}$ is 
the Fock space of the second-quantized 
D5-brane which allows 
the $U(n)$ Chan-Paton factors.}  
${\cal O}_m$ will be identified with 
a marginally stable bound state of 
$m$ D5-branes. 
This identification seems to be reasonable since 
${\cal O}_m$ will be regarded 
to be constructed from the local data of the resolution 
of the singularity caused by overlapping 
$m$ D5-branes. 
We may introduce 
the creation and annihilation operators 
of these bound states. 
The creation and annihilation opeartors 
of ${\cal O}_m$ are denoted 
respectively by $\alpha_{-m}$ and $\alpha_{m}$. 
These operators can be 
normalized to satisfy the relations 
\begin{eqnarray} 
\mbox{$[~\alpha_{m_1}~,~\alpha_{m_2}~]$} 
=\delta_{m_1+m_2,0}~~. 
\label{CCR1}
\end{eqnarray}
Notice that they are bosonic operators 
since the degrees of ${\cal O}_m$, 
which count the fermionic contributions  
to ${\cal O}_m$ 
\footnote{From the topological field theoretical 
viewpoint the degrees of ${\cal O}_m$ 
can be interpreted as the ghost numbers
.}, 
are even.

\newpage 
\subsection*{Acknowledgements}

        We would like to thank H.Nakajima for sending us 
his beautiful lecture note \cite{Nakajima-Lec}. 
We benefited from discussions with H.Kunitomo and K.Furuuchi.  
We also thank J.A.Harvey for letting us know the paper 
\cite{Harvey-Moore} in which Nakajima's results are nicely 
reviewed from the viewpoint of string dualities.

\newpage

\subsection*{Table 1~~~$(\hat{B}_a,\hat{H}_{\dot{A}})$}

\begin{eqnarray} 
&& 
\nonumber \\
\hat{B}_{1} 
&=&
\left( 
\renewcommand{\arraystretch}{0.5} 
\begin{array}{ccccc|ccccccccc}
0 & 1 & 0 & \ldots & 0 & 
0 & 0 & \ldots & \ldots & \ldots & \ldots & \ldots & \ldots & 0 
\\  
0 & 0 & 1 & \ddots & \vdots &    
\vdots & \vdots & & & & & & & \vdots 
\\
\vdots & \ddots & \ddots & \ddots & 0 &    
0 & 0 & \ldots & \ldots & \ldots & \ldots & \ldots & \ldots & 0 
\\ 
\vdots & & \ddots & \ddots & 1 & 
0 & 0 & \ldots & \ldots & \ldots & \ldots & \ldots & \ldots & 0  
\\ 
0 & \ldots & \ldots & 0 & 0 &  
0 & 0 & \ldots & \ldots & \ldots & \ldots & \ldots & \ldots & 0 
\\ 
\hline
0 & \ldots & \ldots & \ldots & 0 &     
0 & 1 & 0 & \ldots & \ldots & \ldots & \ldots & \ldots & 0  
\\  
0 & \ldots & \ldots & \ldots & 0 &  
0 & 0 & 1 & 0 & \ldots & \ldots & \ldots & \ldots & 0  
\\
\vdots & & & & \vdots & 
\vdots & \vdots & \ddots & \ddots & \ddots & & & & \vdots  
\\ 
\vdots & & & & \vdots &  
\vdots & \vdots & & \ddots & \ddots & \ddots & & & \vdots 
\\
\vdots & & & & \vdots &  
\vdots & \vdots & & & \ddots & \ddots & \ddots & & \vdots 
\\ 
\vdots & & & & \vdots &  
\vdots & \vdots & & & & \ddots & \ddots & \ddots & \vdots 
\\ 
\vdots & & & & \vdots &    
\vdots & \vdots & & & & & \ddots & \ddots & 0  
\\
\vdots & & & & \vdots &   
\vdots & \vdots & & & & & & \ddots & 1  
\\
0 & \ldots & \ldots & \ldots & 0 &  
0 & 0 & \ldots & \ldots & \ldots & \ldots & \ldots & \ldots & 0 
\end{array} \right)~~~
\nonumber \\ 
&& 
\nonumber \\
&& 
\nonumber \\
&& 
\nonumber \\ 
\hat{B}_2 &=&
\left( 
\renewcommand{\arraystretch}{0.5} 
\renewcommand{\arraycolsep}{1mm} 
\begin{array}{ccccc|ccccccccc}
\lambda  & \hat{b}_1 & \ldots & \hat{b}_{m\!-\!2} 
& \hat{b}_{m\!-\!1} &  
0 & 0 & \ldots & \ldots & \ldots & \ldots 
& \ldots & \ldots & 0 
\\  
0 & \lambda & \hat{b}_1 & \ldots & \hat{b}_{m\!-\!2} & 
0 & 0 & \ldots & \ldots & \ldots & \ldots 
& \ldots & \ldots & 0 
\\
\vdots  & \ddots  & \ddots & \ddots & \vdots &  
\vdots &  \vdots & & & & & & & \vdots \\ 
\vdots  &  & \ddots & \ddots & \hat{b}_1 &  
\vdots & \vdots & & & & & & & \vdots  
\\ 
0 & \ldots & \ldots & 0 & \lambda  &  
0 & 0 & \ldots & \ldots & \ldots & \ldots 
& \ldots & \ldots & 0 
\\ 
\hline
0 & \ldots & \ldots  & \ldots & 0 & 
0 & \hat{a}_1 & \ldots & \hat{a}_{m'\!-\!1} & \hat{d}_1 
& \ldots & \ldots &\hat{d}_{m\!-\!1}& \hat{d}_m 
\\   
0 & \ldots & \ldots & \ldots & 0 &  
0 & 0 & \hat{a}_1 & \ldots & \hat{a}_{m'\!-\!1} 
& \hat{d}_1 & \ldots & \ldots & \hat{d}_{m\!-\!1}  \\
\vdots & & & & \vdots & 
\vdots & \vdots & \ddots & \ddots & & \ddots & \ddots & & 
\vdots  \\ 
\vdots & & & & \vdots &  
\vdots & \vdots & & \ddots & \ddots & & \ddots & \ddots & 
\vdots  \\
\vdots & & & & \vdots &  
\vdots & & & & \ddots & \ddots & & \ddots & 
 \hat{d}_1 \\ 
\vdots & & & & \vdots &  
\vdots & \vdots & & & & \ddots & \ddots & 
& \hat{a}_{m'\!-\!1} \\ 
\vdots & & & & \vdots &  
\vdots & \vdots & & & & & \ddots & \ddots & \vdots \\
\vdots & & & & \vdots &  
\vdots & \vdots & & & & & & \ddots & \hat{a}_1  \\
0 & \ldots & \ldots & \ldots & 0 &  
0 & 0 & \ldots & \ldots & \ldots & \ldots 
& \ldots & \ldots & 0 
\end{array} \right)~~~ 
\nonumber \\ 
&& 
\nonumber \\
&& 
\nonumber \\ 
&& 
\nonumber \\
\hat{H}_{\dot{1}} 
&=&
\left( 
\renewcommand{\arraystretch}{0.5} 
\begin{array}{cccc} 
0 & \hat{h}_{11} & \ldots & \hat{h}_{1 n\!-\!1} 
\\
\vdots & \vdots & & \vdots 
\\ 
\vdots & \vdots & & \vdots 
\\ 
0 & \hat{h}_{m\!-\!11} & \ldots 
& \hat{h}_{m\!-\!1 n\!-\!1} 
\\
1 & \hat{h}_{m1} & \ldots & \hat{h}_{m n\!-\!1} 
\\ 
\hline 
0 & \hat{h}_{m\!+\!11} & \ldots & 
\hat{h}_{m\!+\!1 n\!-\!1} 
\\ 
\vdots & \vdots & & \vdots 
\\ 
\vdots & \vdots & & \vdots 
\\ 
\vdots & \vdots & & \vdots 
\\ 
0 & \hat{h}_{2\!m\!+\!m'\!-\!1 1} & \ldots & 
\hat{h}_{2\!m\!+\!m'\!-\!1 n\!-\!2}  \\ 
1 & \hat{h}_{2\!m\!+\!m' 1} & \ldots & 
\hat{h}_{2\!m\!+\!m' n\!-\!1}
\end{array} \right)~~~~~~~~~~~~~
\hat{H}_{\dot{2}} ~=~0 
\nonumber 
\end{eqnarray}

\newpage

\subsection*{Table 2~~~$(B_a,H_{\dot{A}})$}

\begin{eqnarray} 
&& 
\nonumber \\
B_1 &=& 
\left( 
\renewcommand{\arraystretch}{0.5} 
\begin{array}{ccccc|ccccccccc}
0 & 1 & 0 & \ldots & 0 & 
0 & 0 & \ldots & \ldots & \ldots & \ldots & \ldots & \ldots & 0 
\\  
0 & 0 & 1 & \ddots & \vdots &    
\vdots & \vdots & & & & & & & \vdots 
\\
\vdots & \ddots & \ddots & \ddots & 0 &    
0 & 0 & \ldots & \ldots & \ldots & \ldots & \ldots & \ldots & 0 
\\ 
\vdots & & \ddots & \ddots & 1 & 
0 & 0 & \ldots & \ldots & \ldots & \ldots & \ldots & \ldots & 0  
\\ 
0 & \ldots & \ldots & 0 & 0 &  
0 & 0 & \ldots & \ldots & \ldots & \ldots & \ldots & \ldots & 0 
\\ 
\hline
0 & \ldots & \ldots & \ldots & 0 &     
0 & 1 & 0 & \ldots & \ldots & \ldots & \ldots & \ldots & 0  
\\  
0 & \ldots & \ldots & \ldots & 0 &  
0 & 0 & 1 & 0 & \ldots & \ldots & \ldots & \ldots & 0  
\\
\vdots & & & & \vdots & 
\vdots & \vdots & \ddots & \ddots & \ddots & & & & \vdots  
\\ 
\vdots & & & & \vdots &  
\vdots & \vdots & & \ddots & \ddots & \ddots & & & \vdots 
\\
\vdots & & & & \vdots &  
\vdots & \vdots & & & \ddots & \ddots & \ddots & & \vdots 
\\ 
\vdots & & & & \vdots &  
\vdots & \vdots & & & & \ddots & \ddots & \ddots & \vdots 
\\ 
\vdots & & & & \vdots &    
\vdots & \vdots & & & & & \ddots & \ddots & 0  
\\
\vdots & & & & \vdots &   
\vdots & \vdots & & & & & & \ddots & 1  
\\
0 & \ldots & \ldots & \ldots & 0 &  
0 & 0 & \ldots & \ldots & \ldots & \ldots & \ldots & \ldots & 0 
\end{array} \right)~~~
\nonumber \\ 
&& 
\nonumber \\ 
&& 
\nonumber \\
&& 
\nonumber \\
B_2 &=& 
\left( 
\renewcommand{\arraystretch}{0.5} 
\renewcommand{\arraycolsep}{1mm} 
\begin{array}{ccccc|ccccccccc}
0 & b_1 & \ldots & b_{m\!-\!2} & b_{m\!-\!1} &  
0 & \ldots & \ldots & 0 & 1 & 0 & \ldots & \ldots & 0 
\\  
0 & 0 & b_1 & \ldots & b_{m\!-\!2} & 
0 & \ldots & \ldots & 0 & 0 & 1 & 0 & \ldots & 0 
\\
\vdots & \vdots & \ddots & \ddots & \vdots &   
\vdots & & & \vdots & \vdots & \ddots & \ddots & \ddots 
& \vdots 
\\ 
\vdots & \vdots & & \ddots & b_1 & 
\vdots & & & \vdots & \vdots & & \ddots & \ddots & 0 
\\ 
0 & 0 & \ldots & \ldots & 0 & 
0 & \ldots & \ldots & 0 & 0 & \ldots & \ldots & 0 & 1 
\\ 
\hline 
c_1 & c_2 & \ldots & c_{m\!-\!1} & c_{m} &    
0 & a_1 & \ldots & a_{m'\!-\!1} & 0 & \ldots & 
\ldots & \ldots & 0  
\\  
0 & c_1 & c_2 & \ldots & c_{m\!-\!1} & 
0 & 0 & a_1 & \ldots & a_{m'\!-\!1} & 0 & 
\ldots & \ldots & 0 
\\
\vdots & \ddots & \ddots & \ddots & \vdots &  
\vdots & \vdots & \ddots & \ddots & & 
\ddots & \ddots & & \vdots 
\\ 
\vdots & & \ddots & \ddots & c_2 & 
\vdots & \vdots & & \ddots & \ddots & & 
\ddots & \ddots & \vdots  
\\
0 & \ldots & \ldots & 0 & c_1 & 
\vdots & \vdots & & & \ddots & \ddots & & \ddots & 0 
\\ 
0 & \ldots & \ldots & 0 & 0 & 
\vdots & \vdots & & & & \ddots & \ddots & &
 a_{m'\!-\!1} 
\\ 
\vdots & & & \vdots & \vdots &     
\vdots & \vdots & & & & & \ddots & \ddots & \vdots 
\\
\vdots & & & \vdots & \vdots &   
\vdots & \vdots & & & & & & \ddots & a_1 
\\
0 & \ldots & \ldots & 0 & 0 &  
0 & 0 & \ldots & \ldots & \ldots & \ldots & \ldots & \ldots & 0 
\end{array} \right) 
\nonumber \\ 
&& 
\nonumber \\ 
&& 
\nonumber \\
&& 
\nonumber \\
H_{\dot{1}} 
&=& 
\left( 
\renewcommand{\arraystretch}{0.5} 
\begin{array}{cccc} 
0 & h_{11} & \ldots & h_{1 n\!-\!1} 
\\
\vdots & \vdots & & \vdots 
\\ 
\vdots & \vdots & & \vdots 
\\ 
0 & h_{m\!-\!11} & \ldots & h_{m\!-\!1 n\!-\!1} 
\\
0 & h_{m1} & \ldots & h_{m n\!-\!1} 
\\ 
\hline 
0 & h_{m\!+\!11} & \ldots & h_{m\!+\!1 n\!-\!1} 
\\ 
\vdots & \vdots & & \vdots 
\\ 
\vdots & \vdots & & \vdots 
\\ 
\vdots & \vdots & & \vdots 
\\ 
0 & h_{2\!m\!+\!m'\!-\!1 1} & \ldots & 
h_{2\!m\!+\!m'\!-\!1 n\!-\!2}  \\ 
1 & h_{2\!m\!+\!m' 1} & \ldots & 
h_{2\!m\!+\!m' n\!-\!1}
\end{array} \right)
~~~~~~~~~~~~~
H_{\dot{2}}~=~0  
\nonumber 
\end{eqnarray}

\end{document}